\documentclass[fleqn,10pt]{wlscirep}
\usepackage[utf8]{inputenc}
\usepackage[T1]{fontenc}
\usepackage{lineno}

\usepackage{tabularx}
  \newcolumntype{L}{>{\raggedright\arraybackslash}X}

\usepackage[super]{nth}
\usepackage[linesnumbered,ruled,vlined]{algorithm2e}

\SetCommentSty{mycommfont}

\usepackage{amsmath}
\usepackage{nccmath}

\usepackage{subfigure}
\title{TBCOV: Two Billion Multilingual COVID-19 Tweets with Sentiment, Entity, Geo, and Gender Labels}

\author{Muhammad Imran}
\author{Umair Qazi}
\author{Ferda Ofli}
\affil{Qatar Computing Research Institute, Hamad Bin Khalifa University, Doha, 34110, Qatar}

\affil[*]{corresponding author: Muhammad Imran (mimran@hbku.edu.qa)}

\begin{abstract}
The widespread usage of social networks during mass convergence events, such as health emergencies and disease outbreaks, provides instant access to citizen-generated data that carry rich information about public opinions, sentiments, urgent needs, and situational reports. Such information can help authorities understand the emergent situation and react accordingly. Moreover, social media plays a vital role in tackling misinformation and disinformation. This work presents TBCOV, a large-scale Twitter dataset comprising more than two billion multilingual tweets related to the COVID-19 pandemic collected worldwide over a continuous period of more than one year. More importantly, several state-of-the-art deep learning models are used to enrich the data with important attributes, including sentiment labels, named-entities (e.g., mentions of persons, organizations, locations), user types, and gender information. Last but not least, a geotagging method is proposed to assign country, state, county, and city information to tweets, enabling a myriad of data analysis tasks to understand real-world issues. Our sentiment and trend analyses reveal interesting insights and confirm TBCOV's broad coverage of important topics.

\end{abstract}
\begin{document}

\flushbottom
\maketitle

\thispagestyle{empty}



\section*{Background \& Summary}
\label{background}

Social media use during emergencies such as natural or human-induced disasters has become prevalent among the masses~\cite{castillo2016big}. Twitter---a microblogging site---is increasingly used by affected people and humanitarian organizations to share and seek information, express opinions, and provide emotional support during disasters~\cite{fraustino2017social,starbird2010chatter}. Prior studies show that Twitter also provides timely access to health-related data about chronic diseases, outbreaks, and epidemics~\cite{sinnenberg2017twitter,zadeh2019social,broniatowski2013national}. Hence, the number of Twitter datasets pertaining to the COVID-19 pandemic has been increasing constantly.
The efficacy of these datasets for various types of analyses can be measured across three fundamental dimensions: Language, Space, and Time (LST).
That is, datasets covering more languages, broader geographical areas, and longer temporal boundaries are preferred for both longitudinal and cross-sectional studies,  
especially during a global emergency such as the COVID-19 pandemic. 
Moreover, training robust machine learning (ML) and natural language processing (NLP) models as well as building predictive analytical tools require large-scale datasets for better inference and generalization. 
However, existing datasets vary in their LST coverage. Many of them are restricted to a single language, e.g., English~\cite{lamsal2020coronasentiment,lamsal2020coronatweets} or Arabic~\cite{alqurashi2020large,haouari2020arcov}, or confined to specific geographies~\cite{kang2020multiscale,park2021covid}. The data collection period also differs from one dataset to another. Among all, the dataset by Banda and colleagues stands out as one of the largest, long-running collections with 383 million tweets~\cite{banda2020large}, however, only a handful of English keywords (only 10) were used for data collection---a common issue with existing datasets. 

To bridge these gaps, we present TBCOV, a large-scale Twitter dataset comprising \textbf{T}wo \textbf{B}illion multilingual tweets related to the \textbf{COV}ID-19 pandemic. Specifically, TBCOV offers 2,014,792,896 tweets collected using more than 800 multilingual keywords over a 14-month period from February \nth{1}, 2020 till March \nth{31}, 2021. 
These tweets span 67 international languages, posted by 87 million unique users across 218 countries worldwide. 
More importantly, covering public chatter on various societal, health, and economic issues caused by the pandemic, TBCOV captures different perspectives and opinions about governments' policy decisions ranging from lock downs to aid allocation for individuals and businesses. It also contains several important implications of the pandemic such as food scarcity, shortage of equipment and supplies, reports of anxiety and depression symptoms, among others.
%
Besides its broad topical and LST coverage, TBCOV is also enriched with several attributes derived from tweet text and meta-data using ML techniques. 
These attributes include sentiment labels, geolocation information, named-entities as well as user types and gender. 
\smallskip

\noindent\textit{Public sentiment:} Distilling tweets to understand people's opinions, emotions, and attitudes towards an issue (e.g., low vaccination rate) or a policy decision has paramount importance for various government entities~\cite{gohil2018sentiment}. Uncertainties in authorities' perception of public sentiment during health crises can otherwise result in poor risk communication~\cite{gui2017managing}. Computational techniques such as sentiment analysis can help authorities to understand aggregated public opinion during crises and devise appropriate strategies~\cite{alamoodi2020sentiment}. To this end, we employ a multilingual transformer-based deep learning model~\cite{barbieri2021xlmtwitter} to tag each tweet in TBCOV according to its sentiment polarity (i.e., \textit{positive}, \textit{neutral}, \textit{negative}).
\smallskip

\noindent\textit{Geolocation information:} Geotagging~\cite{geotagging} is indispensable for geographic information systems (GIS) for timely and effective monitoring of outbreaks, hot-spot prediction, disease spread monitoring, and predictive risk mapping~\cite{boulos2020geographical,haworth2016emergency}. 
User-generated data on social media platforms can fuel many of these applications~\cite{zadeh2019social,tzavella2018opportunities}. However, low prevalence of geo-referenced information on social media poses a challenge. To tackle this challenge, we propose a geolocation tagging approach to map each tweet in TBCOV at \textit{country}, \textit{state}, \textit{county}, or \textit{city} level. 
\smallskip

\noindent\textit{Named-entities:} Mentions of persons, organizations, and locations hold key information in text documents and are crucial for various NLP tasks such as question answering, online reputation management, and automatic text summarization~\cite{marrero2013named}. Named-entity recognition (NER) is a widely used NLP technique to identify references to entities in text documents~\cite{sekine2009named}. Past studies propose several NER techniques ranging from rule- and ML-based to hybrid methods~\cite{farmakiotou2000rule, finkel2009nested}. To identify named-entities in our multilingual tweets, we use language-specific NER models for the most prevalent five languages in TBCOV and one multilingual NER model for all other languages.
\smallskip

\noindent\textit{User types \& gender:} Understanding gender differences is important for addressing societal challenges such as identifying knowledge gaps~\cite{manierre2015gaps}, digital divide~\cite{antonio2014gender}, and health-related issues~\cite{johnson2009better,lawrence2007methodologic}. Tweets in TBCOV are mainly sourced from three types of users, i.e., \textit{individuals}, \textit{organizations}, and \textit{bots}. We first distinguish the user type by running an NER model on the \textit{name} field in a Twitter profile, and then, determine the gender information using an ML classifier if the predicted user type is \textit{individuals}.

To the best of our knowledge, TBCOV is the largest Twitter dataset related to COVID-19 with broad LST coverage and rich information derived from multilingual tweets that can be used for many NLP, data mining, and real-world applications. The dataset is accessible at: \url{https://crisisnlp.qcri.org/tbcov}


\section*{Methods}
\label{sec:methods}
This section summarizes data collection procedures and provides statistics about the dataset. Moreover, it elaborates on the computational techniques employed to derive various attributes such as sentiment labels from tweets.

\subsection*{Data collection and description}
\label{sec:data_colection}
Twitter offers different APIs for data collection. We use the Twitter Streaming API, which allows for collecting tweets based on hashtags/keywords or geographical bounding boxes. 
Following the keyword-based strategy, we started our data collection on February \nth{1}, 2020 using trending hashtags such as \#covid19, \#coronavirus, \#covid\_19, and added new trending hashtags and keywords in the later days. In total, more than 800 multilingual keywords and hashtags encompassing a large set of topics including social distancing, shortages of masks, personal protective equipment (PPE), food, medicine, and reports of COVID-19 symptoms, deaths, were used. Table~\ref{tab:collection_terms} lists some of the terms (full list of keywords is provided in the data release). Twitter offers filtered streams for specific languages; however, we did not filter by any language, and hence, our data is multilingual. Although the data collection was still running at the time of writing this manuscript, all the statistics and analyses presented in this study are based on data collected till March \nth{31}, 2021---i.e., 2,014,792,896 tweets. To the best of our knowledge, this is the largest multilingual Twitter dataset covering a broad spectrum of topics and issues the world has been facing amidst the COVID-19 pandemic. 

\begin{table}[!ht]
\centering
\begin{tabularx}{\linewidth}{L} 
\toprule
\footnotesize
Argentina Coronavirus, Armenia Coronavirus, Australia Coronavirus, Austria Coronavirus, Azerbaijan Coronavirus, Bahamas Coronavirus, Bahrain Coronavirus, Bangladesh Coronavirus, Barbados Coronavirus, Belarus Coronavirus, Belgium Coronavirus, Belize Coronavirus, Benin Coronavirus, Bhutan Coronavirus, Bolivia Coronavirus, Bosnia Herzegovina Coronavirus, Botswana Coronavirus, Brazil Coronavirus, Brunei Coronavirus, Bulgaria Coronavirus, Burkina Coronavirus, Burundi Coronavirus, Cambodia Coronavirus, Cameroon Coronavirus, Canada Coronavirus, COVID-19, Congo COVID-19, Congo COVID-19, Costa Rica COVID-19, Croatia COVID-19, Cuba COVID-19, Cyprus COVID-19, Czech Republic COVID-19, Denmark COVID-19, Djibouti COVID-19, Dominica COVID-19, Dominican Republic COVID-19, East Timor COVID-19, Ecuador COVID-19, Egypt COVID-19, El Salvador COVID-19, Equatorial Guinea COVID-19, Eritrea COVID-19, Estonia COVID-19, Ethiopia COVID-19, Fiji COVID-19, Finland COVID-19, France COVID-19, Gabon COVID-19, Gambia COVID-19, Georgia COVID-19, Germany COVID-19, Ghana COVID-19, \#socialdistancing us, \#socialdistancing usa, \#socialdistancing Alabama, \#socialdistancing Alaska, \#socialdistancing Arizona, \#socialdistancing Arkansas, \#socialdistancing California, \#socialdistancing Colorado, \#socialdistancing Connecticut, \#socialdistancing Delaware, \#socialdistancing Florida, \#socialdistancing Georgia, \#socialdistancing Hawaii, \#socialdistancing Idaho, \#socialdistancing Illinois, \#socialdistancing Indiana, \#socialdistancing Iowa, \#socialdistancing Kansas, \#socialdistancing Kentucky, \#socialdistancing Louisiana, \#socialdistancing Maine, \#socialdistancing Maryland, \#socialdistancing Massachusetts, \#socialdistancing Michigan, económica, quédate en casa Colombia, respiradores Colombia, tapabocas Colombia, UCI disponibles, recuperados covid19 Colombia, muertes Colombia, Nariño Coronavirus, Nariño Covid19, \#coronavirus, \#Corona, \#COVID19, \#WuhanCoronavirus, \#ncoV2019, \#coronavirus, Italia, lombardia, \#covid19italia, \#COVID19Pandemic, Covid, \#CoronavirusAustralia, \#pandemic, Covid-19 USA \\

\bottomrule
\end{tabularx}
\caption{A sample of keywords/hashtags used for data collection}
\label{tab:collection_terms}
\end{table}

\begin{figure}[!t]
\centering
\includegraphics[width=\textwidth]{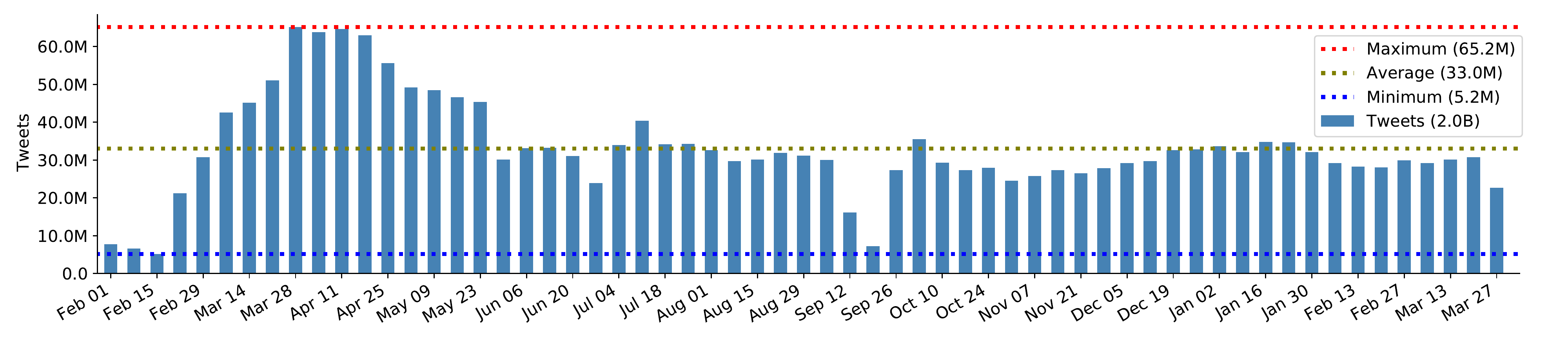}
\caption{Weekly distribution of 2,014,792,896 tweets from February \nth{1}, 2020 to March \nth{31}, 2021.}
\label{fig:data_daily_dist}
\end{figure}

Figure~\ref{fig:data_daily_dist} depicts the volume of tweets ingested across the 61 weeks of the data collection (February \nth{1}, 2020 to March \nth{31}, 2021). The data does not show any gaps, which is an important factor for many types of analysis. The volume of tweets in the first three weeks is relatively lower, e.g., ${\sim}$5 million daily tweets on average. However, a sudden surge can be noticed starting from week four, which amounts to an overall average of 33 million tweets per week. The maximum number of tweets recorded in a week is 65 million.



\begin{figure}[!h]
\centering
\includegraphics[width=\textwidth]{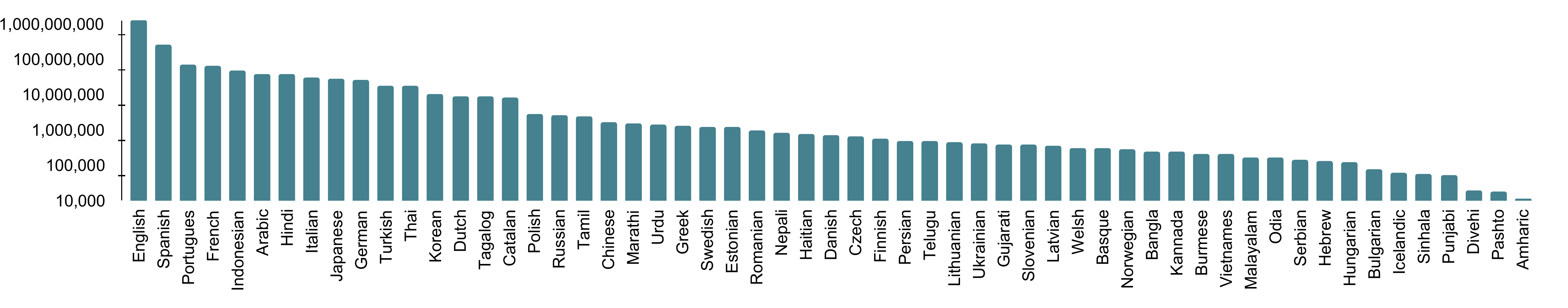}
\caption{Distribution of languages with more than 10K tweets. Number of tweets on y-axis (log-scale).}
\label{fig:lang_dist}
\end{figure}

The tweets in TBCOV dataset are posted by 87,771,834  unique users and among them 268,642 are verified users. 
%
%
In total, the dataset covers 67 international languages. Figure~\ref{fig:lang_dist} shows the distribution of languages (with at least 10K tweets) and the corresponding number of tweets in the log scale. The English language dominates with around 1 billion tweets and the second and third largest languages are Spanish and Portuguese, respectively. There are around 55 million tweets for which the language is undetermined---this is an important set of tweets suitable for the language detection task with code-mixing properties~\cite{thara2018code}. 

The TBCOV dataset is a substantial extension of our previous COVID-19 data release named GeoCoV19~\cite{qazi2020geocov19}. The TBCOV dataset is superior in many ways. First, the TBCOV dataset contains 1.4 billion more tweets than the GeoCoV19 dataset that consists of 524 million tweets. Second, the data collection period of GeoCoV19 was restricted to only four months (Feb 2020 to May 2020), whereas the TBCOV coverage is 14 months (Feb 2020 to March 2021). The third and the most critical extension represents several derived attributes that TBCOV offers, including sentiment labels, named-entities, user types, and gender information. None of these attributes were part of the GeoCoV19 data. Furthermore, the geotagging method used in GeoCoV19 has been substantially improved and used in TBCOV, which yields better inference results.

\subsection*{Named-entity recognition}
\label{sec:entities_extraction}
Named entities represent key elements in a text, including names of \textit{persons}, \textit{organizations}, \textit{locations}, \textit{brands}, and \textit{dates}, among others~\cite{marrero2013named}. Past studies demonstrate diverse applications of named entities on social media such as finding adverse drug reactions~\cite{mackinlay2017detection} and identifying temporal variations of locations, actors, and concepts from tweets about the Zika outbreak~\cite{stefanidis2017zika}. Moreover, such entities, especially in unstructured social media messages, elicit critical information about an event or a situation along different dimensions---\textit{who, where, when, what}---, i.e., people or organizations involved in a situation, their locations, date or time of the event, their tasks, etc. 
Named-entity recognition (NER), i.e., the task of identifying and extracting named entities, serves as the basis of many NLP tasks such as question answering, semantic annotation, information extraction, and text summarization.

\begin{table}[]
\centering
\begin{tabular}{lrrrr}
\toprule
\multicolumn{1}{l}{\textbf{Language}} & \multicolumn{1}{r}{\textbf{Person}} & \multicolumn{1}{r}{\textbf{Organization}} & \multicolumn{1}{r}{\textbf{Location}} & \multicolumn{1}{r}{\textbf{Miscellaneous}}\\
\midrule
English (U) & 14,796,271 & 18,887,285 & 2,930,148 & 10,798,850 \\
English (A) & 409,794,668 & 611,669,779 & 483,680,780 & 1,690,122,455\\
Spanish (U) & 3,777,463 & 2,230,017 & 3,265,204 & 14,968,547\\
Spanish (A) & 98,561,105 & 69,581,078 & 169,903,131 & 301,512,355\\
Portuguese (U) & 1,439,192 & 932,504 & 1,006,396 & 2,845,321 \\
Portuguese (A) & 27,577,759 & 15,896,880 & 40,090,891 & 52,440,351\\
French (U) & 1,374,884 & 804,336 & 719,896 & 3,894,968\\
French (A) & 23,595,420 & 17,256,551 & 34,064,424 & 63,010,283\\
\midrule
Total (U) & \textbf{55,721,884} & \textbf{33,324,173} & \textbf{10,336,415} & \textbf{40,767,983}\\
Total (A) & \textbf{803,832,752} & \textbf{814,205,050} & \textbf{805,175,906} & \textbf{2,320,195,791}\\
\bottomrule
\end{tabular}
\caption{Named-entities extraction results for the top four languages. `U' denotes ``unique occurrences'' and `A' denotes ``all occurrences'' of entities.}
\label{tab:entities_4langs}
\end{table}

Several classical machine learning, and more recently deep learning, techniques have been proposed to perform NER on standard web documents as well as social media data~\cite{li2020survey}. NER techniques and models can be either language-specific (i.e., trained for a particular language) or multilingual (i.e., trained to operate on multiple languages). Language-specific models usually outperform multilingual models. Hence, we use five language-specific, deep learning-based NER models targeting the top five languages in our dataset, i.e., English, Spanish, Portuguese, French, and Italian, and a multilingual deep learning model to cover the remaining 62 languages. The English NER model can recognize eighteen different types of entities, including \textit{persons, organizations, locations, language, product, time, money}, etc. However, all other NER models can detect only the three fundamental entity types, i.e., \textit{persons}, \textit{organizations}, and \textit{locations}, in addition to a \textit{miscellaneous} type representing other entities. We introduced an additional entity, named \textit{covid19}, to represent different COVID-19 related terms ($N=60$), including Coronavirus, SARS-CoV, SARS-COVID-19, Corona, Covid19, etc. All six models and their performance scores are publicly available\footnote{\url{https://spacy.io/}}. Text of all two billion tweets was first preprocessed by removing URLs, usernames, emojis, and other special characters, and then fed to one of the six NER models depending on the \textit{language} attribute. Four NVIDIA Tesla P100 GPUs were used to process all the data. The entities recognition and extraction process resulted 4.7 billion entities from all tweets. Table~\ref{tab:entities_4langs} shows the number of entities extracted of type \textit{person}, \textit{organization}, \textit{location}, and \textit{misc} (i.e., miscellaneous) for the top four languages. The selected languages represent 38\% of \textit{person}, 68\% of \textit{organization}, and 76\% of \textit{location} out of all the extracted entities. The remaining entities represent a long-tail distribution. 
\subsection*{Geographic information}
\label{sec:geo_inferrence}
Geotagged social media messages with situational or actionable information have a profound impact on decision-making processes during emergencies~\cite{GRACE2021101923,zade2018situational}. For example, recurring tweets showing face mask violations in a shopping mall or a park, or on a beach, can potentially inform authorities' decisions regarding stricter measures. Moreover, when governments' official helplines are overwhelmed~\cite{hindustan_times_helpline_21}, social media reports, e.g., shortages of essential equipment in a remote hospital or patients stuck in traffic requiring urgent oxygen supply~\cite{timesofindia_sm_use21}, could be life-saving if processed and geotagged timely and effectively. Furthermore, GIS systems, which heavily rely on geotagged information, are critical for many real-world applications such as mobility analysis, hot-spot prediction, and disease spread monitoring. Despite these advantages, social media messages are often not geotagged, thus not suitable for automatic consumption and processing by GIS systems. However, they may still contain toponyms or place names such as street, road, or city---information useful for geotagging. 

\begin{algorithm}[!ht]
\SetAlgoLined
\DontPrintSemicolon

\SetKwFunction{FMain}{Main}
\SetKwFunction{FGeoLoc}{geoLocalizeText}
\SetKwFunction{FMajority}{getMajorityLocation}
\SetKwFunction{FSub}{Sub}
$NER\_models \gets getNERModels(lang)$ \tcp{load all six NER models into a dictionary where the lang parameter indicates languages, i.e., en, fr, es, pt, it, and ml}
\SetKwProg{Fn}{Function}{:}{}
  \Fn{\FGeoLoc{$input\_text, lang$}}{
        Initialize $address\_objects[~]$  \tcp{initializing address objects array to store Nominatim responses}
        Initialize $entities[~]$ \tcp{initializing entities array to store NER model responses}
        $processed\_text \gets preprocessing(input\_text)$ \tcp{remove URLs, emoticons, usernames, special characters}
        $entities \gets NER\_models[lang].getEntities(processed\_text)$\\
       \For{$idx, entity\;$in$\; entities$}
        {
            \If{$entity = LOC\; or\; FAC\; or\; GPE$}
            {
            $address \gets geocoding(entity)$ \tcp{nominatim server call}
            $address\_objects[idx] \gets address$
            }
       }
        \KwRet $address\_objects$
  }
 \caption{Pseudo-code for processing toponyms from text}
 \label{algo:geo_from_text}
\end{algorithm}

\begin{algorithm}[!ht]
\SetAlgoLined
\DontPrintSemicolon
\SetKwFunction{FMain}{Main}
\SetKwFunction{FGeoLoc}{localizeText}
\SetKwFunction{FMajority}{getMajorityLocation}
\SetKwFunction{FPlace}{geoLocalizePlace}

\SetKwProg{Fn}{Function}{:}{}
  \Fn{\FPlace{$place\_object$}}{
        \uIf{$place\_object[``place\_type''] = POI$} 
        {
            $coordinates \gets place\_object[``place''][``coordinates'']$\;
            $address \gets reverseGeocoding(coordinates)$ \tcp{Nominatim server call}
        }
        \uElseIf{$place\_object[``place\_type''] = city \; OR \; neighborhood \; OR\; admin \;OR\; country$}
        {
            \If {$place\_object[``full\_name'']\; is \;not \;None$}
            {
                $address \gets geocoding(``full\_name'')$ \tcp{Nominatim server call}
                $granularity\_level = getLowestGranularity(place\_object[``place\_type''])$\;
                \If{$address[``country\_name''] != place\_object[``country\_name''] \;AND\; granularity\_level\; != ``Country''$}
                {
                     $bounding\_box \gets place\_object[``place''][``coordinates'']$ \;
                     $longitude, latitude \gets getMidpoint(bounding\_box)$ \tcp{Get the midpoint (latitude and longitude) of the bounding box}
                     $address \gets reverseGeocoding(longitude, latitude)$ \tcp{Nominatim server call}
                }
            }
            \ElseIf {$place\_object[``name'']\; is \;not \;None$}
            {
                $address \gets geocoding(``name'')$ \tcp{Nominatim server call}
                $granularity\_level = getLowestGranularity(place\_object[``place\_type''])$\;
                \If{$address[``country\_name''] != place\_object[``country\_name''] \;AND\; granularity\_level\; != ``Country''$}
                {
                     $bounding\_box \gets place\_object[``place''][``coordinates'']$ \;
                     $longitude, latitude \gets getMidpoint(bounding\_box)$ \tcp{Get the midpoint (latitude and longitude) of the bounding box}
                     $address \gets reverseGeocoding(longitude, latitude)$ \tcp{Nominatim server call}
                }
            }
        }
        \KwRet $address$
  }
 \caption{Pseudo-code for geotagging place object}
 \label{algo:geo_from_place}
\end{algorithm}

\subsubsection*{Geotagging approach}
This work geotags tweets using five meta-data attributes. Three of them, i.e., \textit{tweet text}, \textit{user location}, and \textit{user profile description}, are free-form text fields potentially containing toponym mentions. The \textit{tweet text} attribute, which represents the actual content of a tweet in 280 characters, can have multiple toponym mentions for various reasons. 
The \textit{user location} is an optional field that allows users to add location information such as their country, state, and city whereas the \textit{user profile description} field usually carries users' demographic data~\cite{sloan2015tweets}. The latter two user-related attributes are potential sources for user location inference~\cite{ajao2015survey}. The remaining two attributes, i.e., \textit{geo-coordinates} and \textit{place tags} carry geo-information in a structured form that is suitable for the direct consumption by the automatic GIS systems. The \textit{geo-coordinates} field contains \textit{latitude} and \textit{longitude}, which are directly obtained from the users' GPS-enabled devices. However, many users refrain from enabling this feature, thus only 1-2\% of tweets contain exact coordinates~\cite{carley2016crowd}. The \textit{place} attribute carries a bounding box representing a location tag that users optionally provide while posting tweets. Although \textit{geo-coordinates} and \textit{place} attributes suit GIS consumption, for the sake of standardization with text-based attributes, we convert them to country, state, county, and city-level information using a process known as \textit{``reverse geocoding''} which is described next.

The \textit{pseudo-codes} of the proposed geotagging procedures are presented in Algorithms \ref{algo:geo_from_text},~\ref{algo:geo_from_place}, \& \ref{algo:geo_main}. Two common processes across three procedures are (i) \textit{geocoding} and (ii) \textit{reverse geocoding}. The \textit{geocoding} process is used to obtain geo-coordinates from a given place name (e.g., California) while the \textit{reverse geocoding} process is used to retrieve the place name corresponding to a given geo-coordinates. Multiple geographical databases exist and support these two processes. We use the Nominatim database\footnote{https://nominatim.org/}, which is a search engine of OpenStreetMap\footnote{https://www.openstreetmap.org/}. The official online Nominatim service restricts 60 calls/minute, and hence, is not suitable for us to make billions of calls in a reasonable time period. Therefore, we set up a local installation of the Nominatim database. Both Nominatim calls (i.e.,  \textit{geocoding} and \textit{reverse geocoding}) return, among others, a dictionary object named \textit{``address''}, which depending on the location granularity, comprising several attributes such as \textit{country}, \textit{state}, \textit{county}, and \textit{city}.

The procedure to process toponyms from text fields (except \textit{user location}) is highlighted in Algorithm \ref{algo:geo_from_text}. The procedure assumes that all six NER models are already loaded (line 1). After initializing the required arrays, preprocessing of the text (i.e., remove all URLs, usernames, emoticons, etc.) is performed (line 3). The \textit{lang} attribute, which represents the language of a tweet, determines the NER model to be applied on the processed text for entity extraction. Recall that five language-specific and one multilingual NER models are used in this study. Since NER models return different types of entities, next we iterate over all predicted entities (line 7) to retain the ones with the following types: \textit{LOC}, \textit{FAC}, or \textit{GPE} (line 8). The \textit{LOC} entity type represents locations, mountain ranges, bodies of water; the \textit{FAC} corresponds to buildings, airports, highways, bridges, etc., and \textit{GPE} represents countries, cities, and states. Finally, a \textit{geocoding} call per entity is made and responses are stored (line 9 \& 10).

Algorithm~\ref{algo:geo_from_place} outlines the procedure for processing the \textit{place} attribute. The \textit{place\_type} attribute inside the \textit{place} object helps determine if a reverse or a simple geocoding call is required (lines 2 \& 5). Places of type \textit{POI} (Point-of-Interest) contain exact \textit{latitude} and \textit{longitude} coordinates, and thus, suitable to perform \textit{reverse geocoding} calls (line 4). However, \textit{non-POI} places (i.e., \textit{city}, \textit{neighborhood}, \textit{admin} or \textit{country}) are represented with a bounding box spanning a few hundred square feet (e.g., for buildings) to thousands of square kilometers (e.g., for cities or countries). Moreover, large bounding boxes can potentially cover multiple geographic areas, e.g., two neighboring countries, and hence, can be ambiguous to resolve. To tackle this issue, we use \textit{full\_name} attribute to make \textit{geocoding} calls (lines 7 \& 16) and compare the \textit{country name} of the obtained \textit{address} with that of the original place object (lines 9 \& 18). In case countries do not match, as a last resort, a midpoint of the bounding box is obtained (lines 11 \& 20) to make \textit{reverse geocoding} calls (lines 12 \& 21).

\begin{algorithm}[!t]
\SetAlgoLined
\DontPrintSemicolon
 $tweets[~] \gets load\_tweets\_batch()$\;
 \For{$tweet~in~tweets$}{
  \If{$tweet[``coordinates"] \; is\; not\; None$}
   {
    $longitude, latitude \gets tweet[``coordinates"][``coordinates"]$\;
   $adrsGeo \gets reverseGeocoding(longitude, latitude)$ \tcp{nominatim server call}
   }
   \If{$tweet[``place"] \; is\; not\; None$}
   {
   $adrsPlace \gets geoLocalizePlace(tweet[``place"])$
   }
   \If{$tweet[``text"] \; is\; not\; None$}
   {
   $adrsText \gets geoLocalizeText(tweet[``full\_text"])$
   }
   \If{$tweet[``user"][``location"] \; is\; not\; None$}
   {
   $processed\_UserLoc \gets preprocessing(tweet[``user"][``location"])$ \tcp{remove URLs, emoticons, usernames, special characters}
   $adrsUserLoc \gets geocoding(processed\_UserLoc)$ \tcp{nominatim server call}
   }
   \If{$tweet[``user\_profile\_description"] \; is\; not\; None$}
   {
   $adrsUserProfile \gets geoLocalizeText(tweet[``user\_profile\_description"])$
   }
   \KwRet $adrsGeo, adrsPlace, adrsText, adrsUserLoc, adrsUserProfile$
 }
 \caption{Pseudo-code for the overall processing of all attributes}
 \label{algo:geo_main}
\end{algorithm}

Algorithm~\ref{algo:geo_main} outlines the \textit{pseudo-code} of the overall geotagging process. It starts with loading a batch 
of tweets (line 1) and iterating over them (line 2). Tweets with \textit{coordinates} are used to make a \textit{reverse geocoding} call (lines 3--5). For \textit{place} tweets, the \textit{geoLocalizePlace} procedure is called, which is defined in Algorithm~\ref{algo:geo_from_place}. And, for the two text-based attributes (i.e., \textit{text}, \textit{user profile description}), the \textit{geoLocalizeText} procedure is called, which is defined in Algorithm~\ref{algo:geo_from_text}. However, the \textit{user location} attribute is pre-processed and geo-coded without applying the NER model (lines 13--15). The evaluation results of the proposed geotagging approach are presented in the next section.

\begin{table}[!h]
\centering
\begin{tabular}{lrrrrrrr}
\toprule
\multicolumn{1}{l}{Location attribute} & \multicolumn{1}{r}{Total occurrences} & \multicolumn{1}{r}{Geotagged (yield \%)} & \multicolumn{1}{r}{Countries} & \multicolumn{1}{r}{States} & \multicolumn{1}{r}{Counties} & \multicolumn{1}{r}{Cities}\\
\midrule
Geo coordinates & 2,799,378 & 2,799,378 (100\%) & 211 & 1,912 & 9,037 & 8,079\\
Place & 51,411,442 & 51,061,938 (99\%) & 215 & 1,906 &13,343 & 9,932\\
User location & 1,284,668,011 & 1,132,595,646 (88\%) & 218 & 2,511 & 24,806 & 20,648\\
User profile description & 1,642,116,879 & 180,508,901 (11\%)& 218 & 2,485 & 18,588 & 14,600\\
Tweet text & 2,014,792,896 & 515,802,081 (26\%)& 218 & 2,513 & 24,235 & 20,549\\
\bottomrule
\end{tabular}
\caption{Geotagging results for all five attributes with unique occurrences, geotagging yield, and resolved countries, states, counties, and cities}
\label{tab:geotagging_results}
\end{table}

\begin{figure}[h]
\centering   
\subfigure[Monthly proportion of tweets from top-10 countries]{\label{fig:top10_countries_dist}\includegraphics[width=0.49\linewidth]{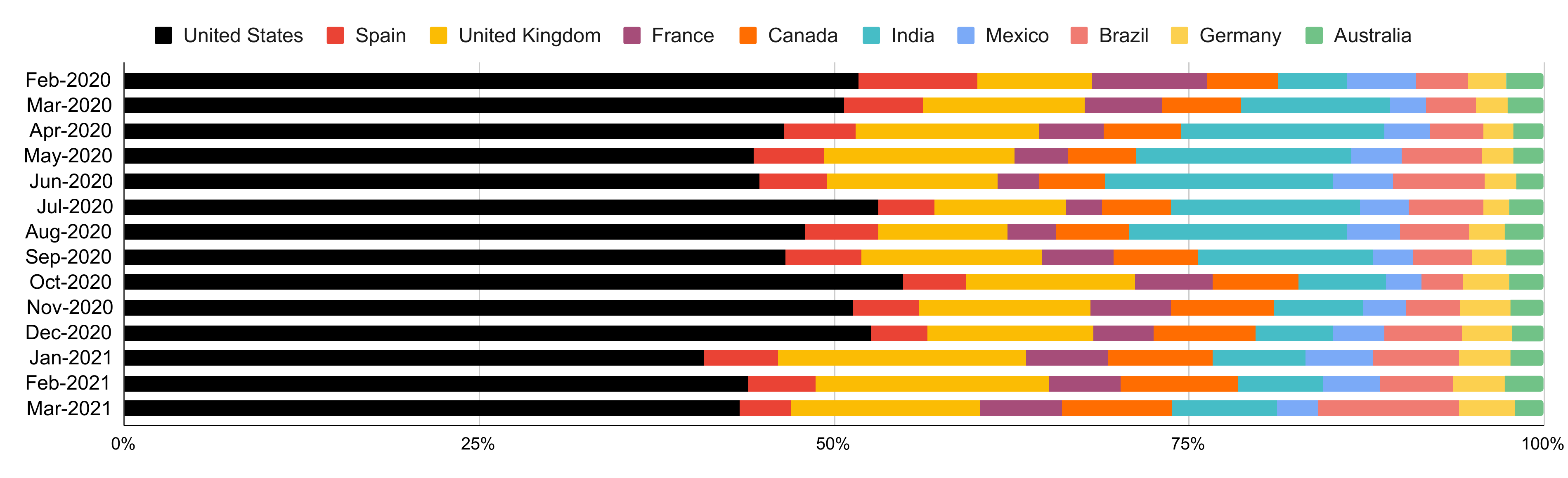}}
\subfigure[Monthly proportion of tweets from top-10 cities]{\label{fig:top10_cities_dist}\includegraphics[width=0.49\linewidth]{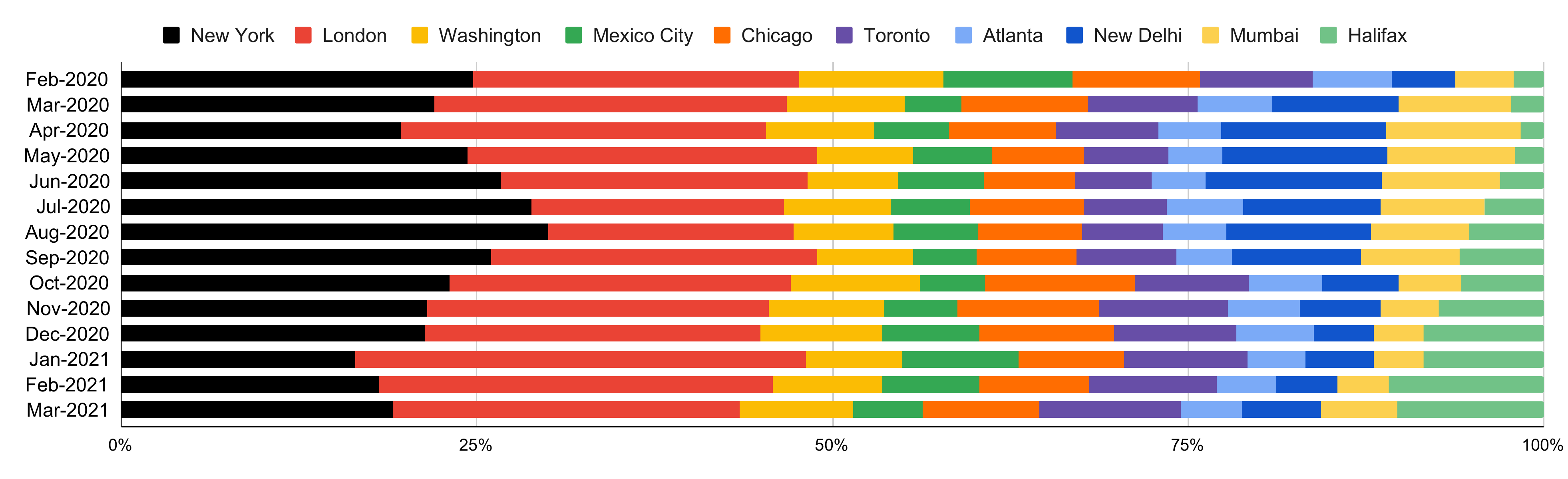}}
\caption{Countries and cities distributions across months sorted by their overall tweets}
\label{fig:top_10_countries_citites}
\end{figure}

The geotagging approach identified 515,802,081 mentions of valid toponyms from \textit{tweet text} and 180,508,901 from \textit{user profile description}. More importantly, out of all 1,284,668,011 users' self-declared locations in the \textit{user location} field, 1,132,595,646 (88\%) were successfully geotagged . Moreover, the process yielded 2,799,378 and 51,061,938 locations for \textit{geo-coordinates} and \textit{place} fields, respectively. Table~\ref{tab:geotagging_results} shows important geotagging results, including total occurrences, geotagging yield, and resultant resolved locations granularity at \textit{country}, \textit{state}, \textit{county}, and \textit{city} level. To determine the country, state, county, and city of a tweet, we mainly rely on three attributes. The first two attributes are users' self-reported location in the \textit{user location} or \textit{user profile description} fields. GPS coordinates are used (if available) in case a tweet is not resolved through \textit{user location} and \textit{user profile description} fields. Altogether, $>$1.8 billion locations corresponding to 218 unique countries, 2,518 sates, 26,605 counties, and 24,424 cities worldwide were resolved. The dataset contains 175 countries and 609 cities around the world having at least 100K tweets. Figure~\ref{fig:top_10_countries_citites} depicts the monthly distribution of top 10 countries and cities throughout the data collection period.

\begin{figure}[!h]
\centering
\includegraphics[width=0.95\textwidth]{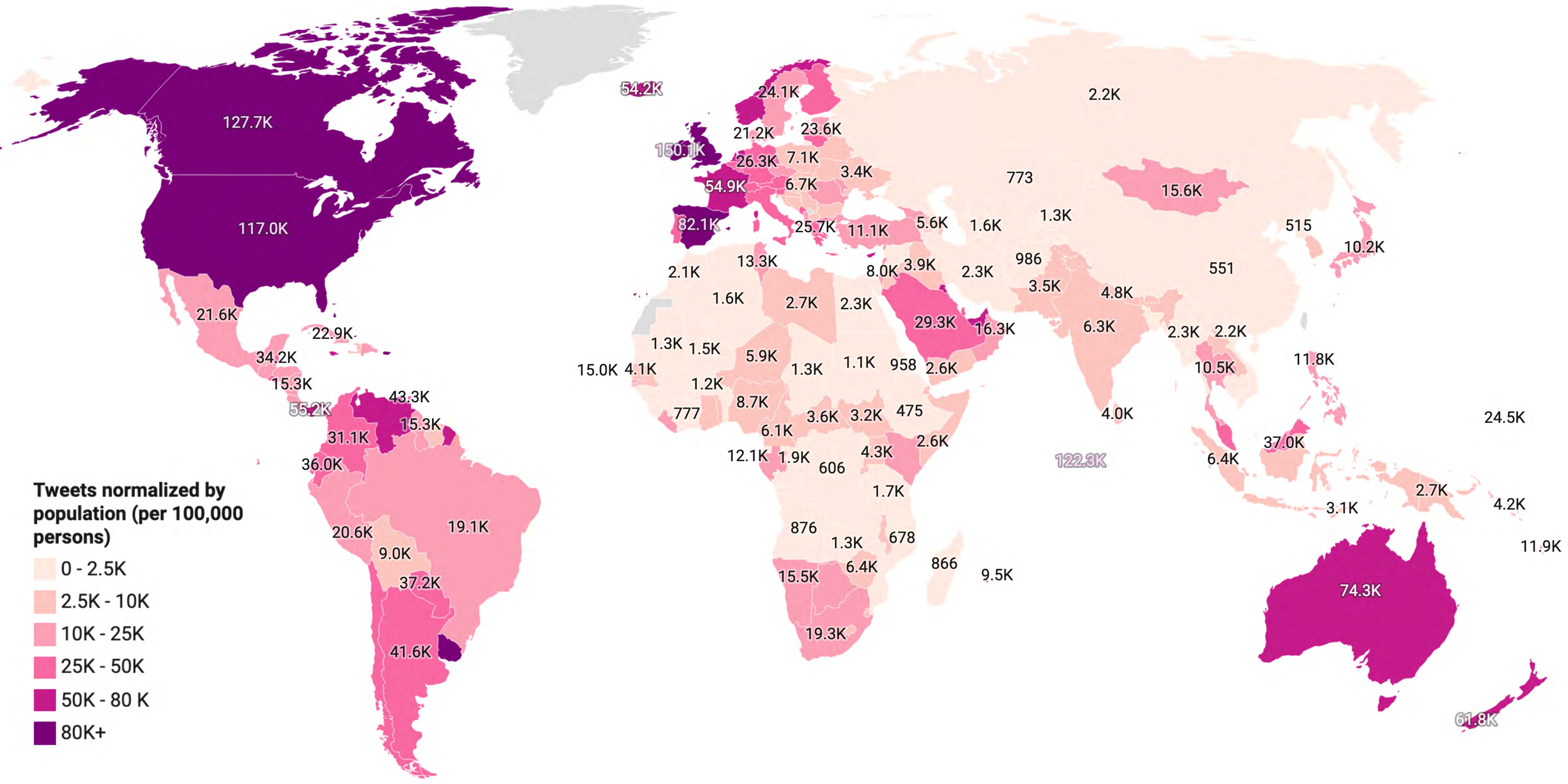}
\caption{Geotagged tweets worldwide normalized by country's population (per 100,000 persons). Tweets geotagged using \textit{user location}, \textit{user profile description}, and \textit{GPS-coordinates} are used.}
\label{fig:worldwide_tweets_normalized}
\end{figure}

To allow meaningful comparisons of geotagged tweets across different countries, we normalize tweets from each country by its population and calculate posts per 100,000 persons. For this purpose, geotagged tweets resolved through \textit{user location}, \textit{user profile description}, and \textit{geo coordinates} attributes were used. Figure~\ref{fig:worldwide_tweets_normalized} shows the normalized counts of geotagged tweets for each country on a world map.

\subsection*{Sentiment classification}
\label{sec:sentiment_classification}
Understanding public opinion and sentiment is important for governments and authorities to maintain social stability during health emergencies and disasters~\cite{huang2021impact,zhang2021temporal}. Prior studies highlighted social networks as a potential medium for analyzing public sentiment and attitude towards a topic~\cite{o2010tweets}. Opinionated messages on social media can vary from reactions on a policy decision\cite{burnap2015cyber} or expressions of sentiment about a situation\cite{beigi2016overview} to sharing opinions during sociopolitical events such as Arab Spring~\cite{aday2012new}. Sentiment analysis, which is a computational method to determine text polarity, is a growing field of research in the text mining and NLP communities~\cite{liu2012sentiment}. There is a vast literature on the algorithms and techniques proposed for sentiment analysis---detailed surveys can be found in~\cite{medhat2014sentiment,zhang2018deep,yue2019survey}. Moreover, numerous studies employ sentiment analysis techniques to comprehend public sentiment during events ranging from elections, sports, to health emergencies~\cite{beigi2016overview,ceron2014every}. We are interested in understanding the public sentiment perceived from multilingual and multi-topic COVID-19 tweets from worldwide.

Our Twitter data is multilingual and covers dozens of real-world problems and incidents such as lockdowns, travel bans, food shortages, among others. Thus, sentiment analysis models that focus on specific topics or domains and support specific languages do not suit our purpose. 
The NLP community offers a myriad of multilingual architectures ranging from LSTMs to more famous transformer-based models~\cite{yue2019survey}. Most recently, a transformer-based model called XLM-T has been proposed as a multilingual variant of the XLM-R model~\cite{conneau2020unsupervised} by fine-tuning it on millions of Twitter general-purpose data in eight languages~\cite{barbieri2021xlmtwitter}. Although the original XLM-R model is trained on one hundred languages using more than two terabytes of filtered CommonCrawl data\footnote{\url{https://commoncrawl.org/}}, its Twitter variant XLM-T achieves better performance on a large multilingual benchmark for sentiment analysis~\cite{barbieri2021xlmtwitter}. We used the XLM-T model to obtain sentiment labels and confidence scores for all two billion tweets in our dataset. Next, we highlight important distributions and present our brief analyses of the obtained sentiment labels. 

\begin{figure}[!t]
\centering
\includegraphics[width=0.8\textwidth]{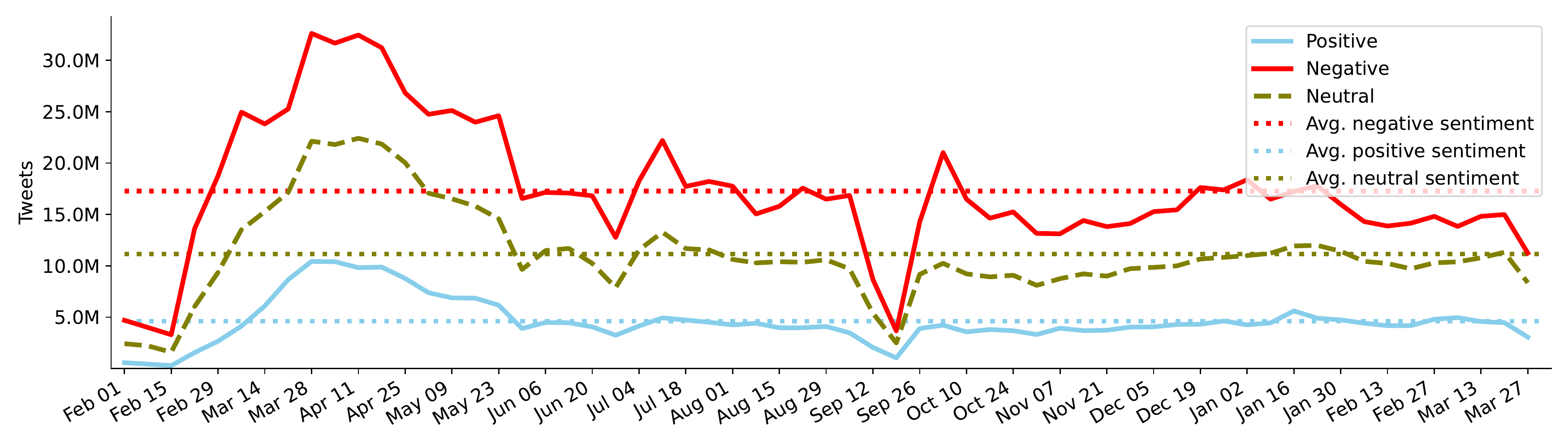}
\caption{Weekly distribution representing public sentiment based on worldwide tweets in all languages.}
\label{fig:world_weekly_sent}
\end{figure}

Of all two billion tweets, 1,054,008,922 (52.31\%) labeled as negative, 680,300,793  (33.77\%) as neutral, and 280,483,181 (13.92\%) as positive. Figure~\ref{fig:world_weekly_sent} presents weekly aggregation of sentiment labels for all tweets in all languages. As anticipated, the negative sentiment dominates throughout (i.e., all 14 months) the data collection period. A significant surge of negative sentiment is apparent in the beginning of March, peaking in April (first week), and then averaging down during the later months. Several hills and valleys appear, but no weeks after April 2020 reaches as high as negative tweets surged in April. The neutral sentiment worldwide stays always lower than the negative, but follows a similar pattern as in the case of the negative sentiment. Not surprisingly though, the positive sentiment remains the lowest sentiment expressed in tweets with steady average except a few weeks in April 2020.

\begin{figure}[!ht]
\centering
\includegraphics[width=\textwidth]{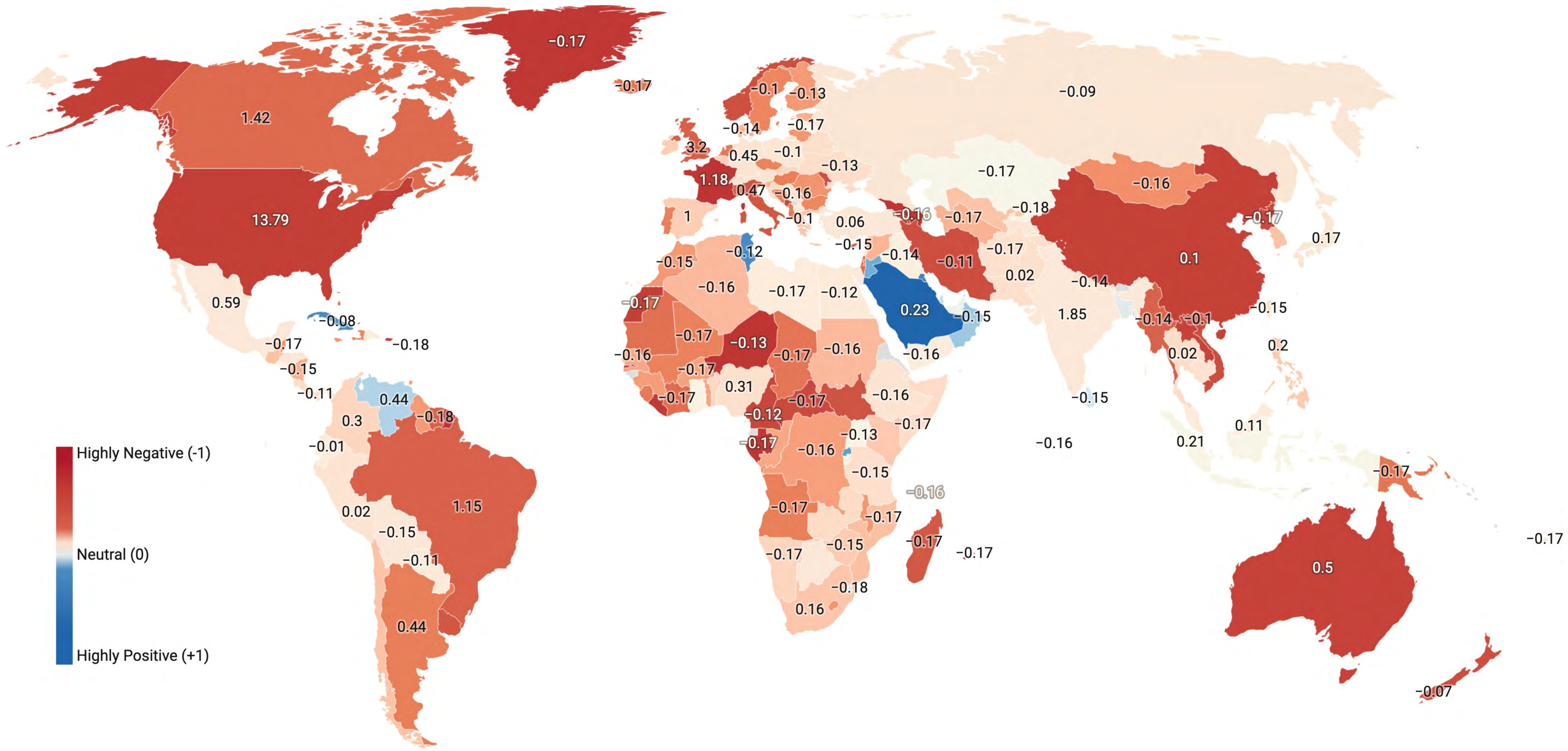}
\caption{Worldwide sentiment based on normalized classifier scores of the representative sentiment in each country. Numbers on countries are z-scores computed using the representative sentiment tweets normalized by total tweets from all countries.}
\label{fig:world_sentiment_map}
\end{figure}

Figure~\ref{fig:world_sentiment_map} shows countries' aggregated sentiment on a world map. The sentiment scores for countries represent normalized weighted averages based on the total number of tweets from a country and model's confidence scores for positive, negative, and neutral tweets. Equation~\ref{eq:avg_sent_score} shows the computation of weighted average sentiment score for a country: 

\begin{ceqn}
\begin{equation}
\label{eq:avg_sent_score}
 S_c = \frac{\sum_{t^{c}_{i} \in \{pos, neut\}}\Theta^{c}_{i} - \sum_{t^{c}_{i} \in \{neg\}}\Theta^{c}_{i}}{N_c} 
\end{equation}
\end{ceqn}

\noindent where $t^{c}_{i}$ represents the sentiment label of tweet $i$ form country $c$ while $\Theta^{c}_{i}$ indicates the model's confidence score for $t^{c}_{i}$, and $N_c$ corresponds to the total number of tweets from the country. The normalized scores range from -1 to 1, where -1 represents high-negative and 1 high-positive, with zero being neutral. The model confidence score represents the model's trust level for assigning a sentiment class to a tweet and it ranges between 0 and 1. The numbers on top of each country are z-scores computed using the representative sentiment tweets normalized by the total tweets from all countries.
Overall, the map shows overwhelming negative sentiment across all except a few countries. Surprisingly, Saudi Arabia and other Gulf countries, including Qatar, UAE, Kuwait, show a strong positive sentiment. Rest of the world, including the US, Canada, and Australia, show moderate to strong negative sentiment.

\begin{figure}[!t]
\centering    
\subfigure[United States]{\label{fig:us_sent}\includegraphics[width=0.49\columnwidth]{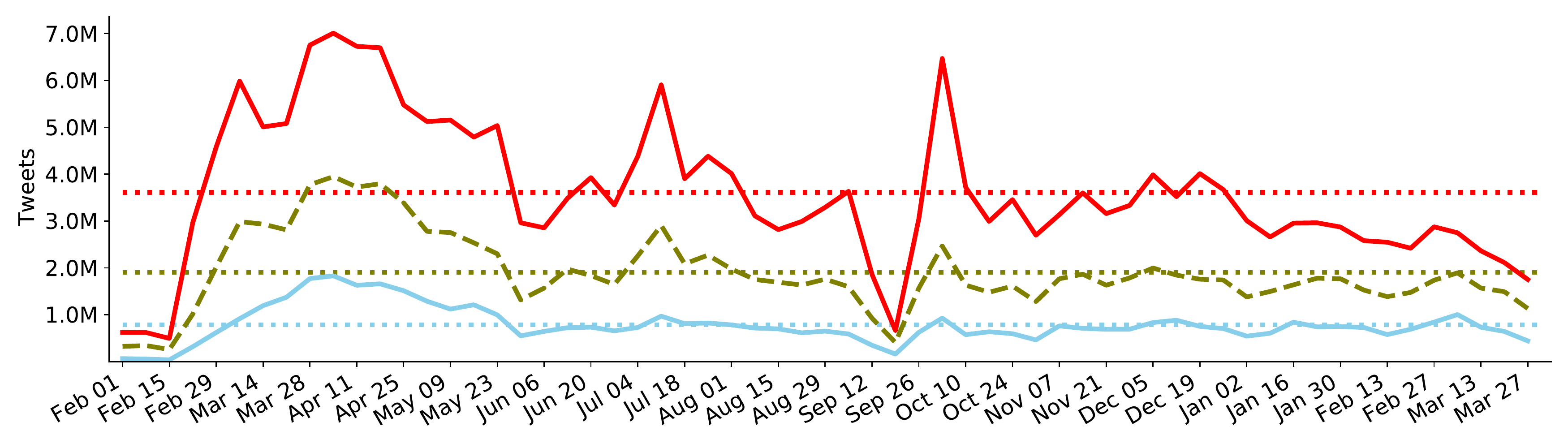}}
\subfigure[United Kingdom]{\label{fig:uk_sent}\includegraphics[width=0.49\columnwidth]{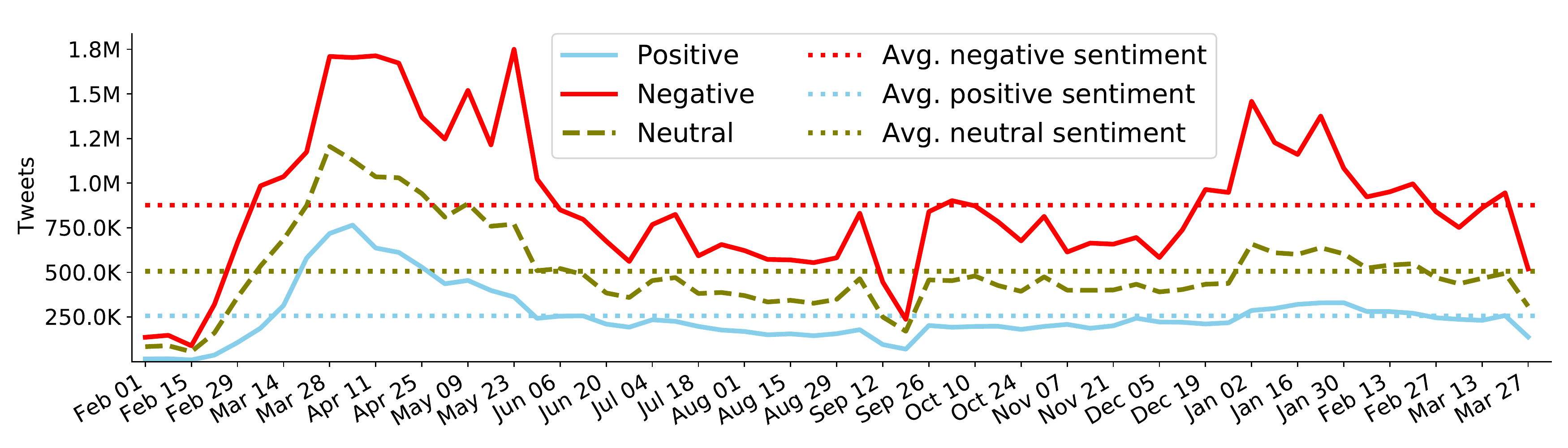}}
\subfigure[India]{\label{fig:india_sent}\includegraphics[width=0.49\columnwidth]{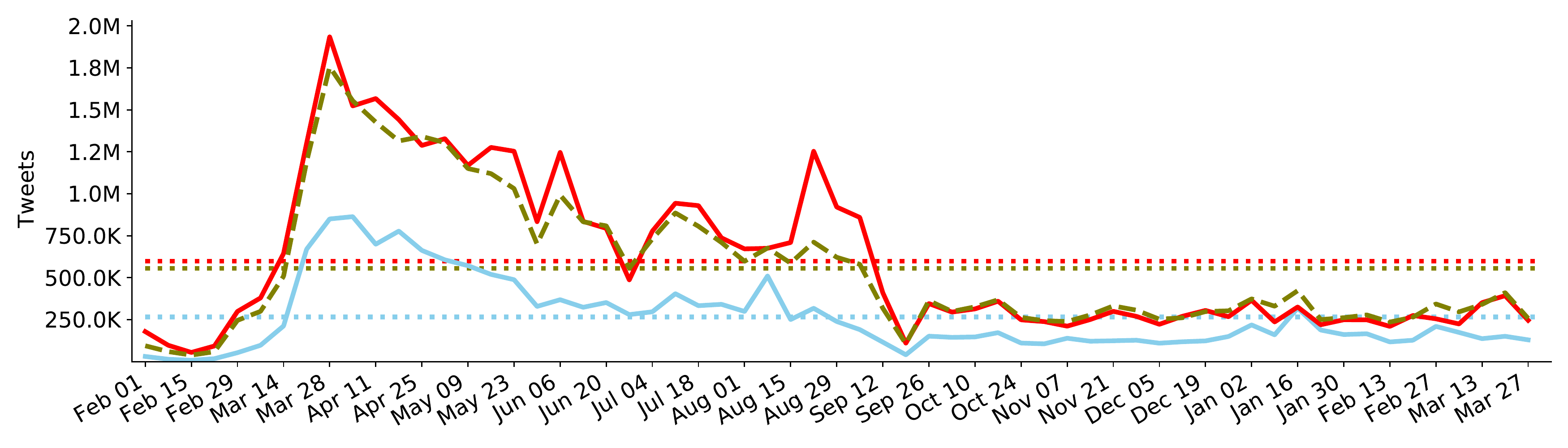}}
\subfigure[Canada]{\label{fig:canada_sent}\includegraphics[width=0.49\columnwidth]{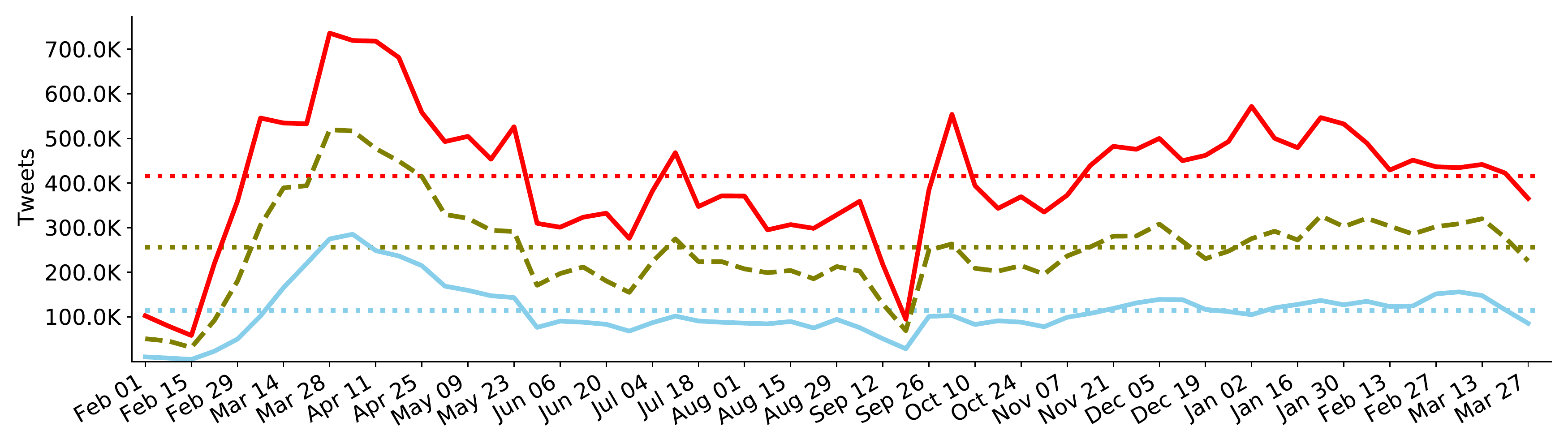}}
\subfigure[Brazil]{\label{fig:brazil_sent}\includegraphics[width=0.49\columnwidth]{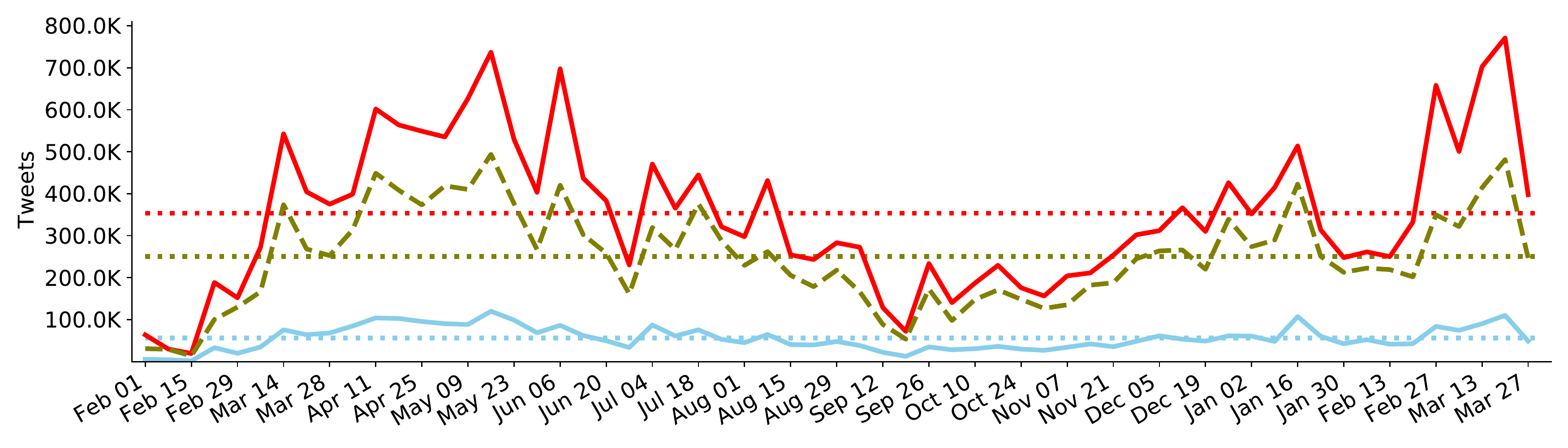}}
\subfigure[Spain]{\label{fig:spain_sent}\includegraphics[width=0.49\columnwidth]{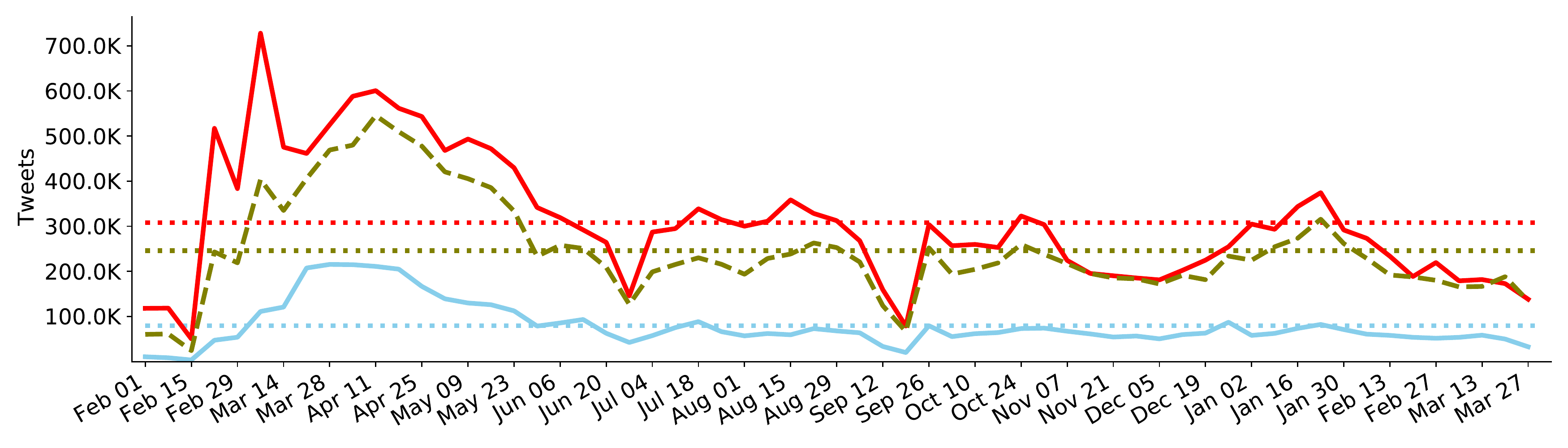}}\vspace{-0.4cm}
\caption{Weekly distribution of sentiment labels for the top six countries}
\label{fig:countries_weekly_sent}
\end{figure}

Figure~\ref{fig:countries_weekly_sent} shows the weekly sentiment trends for the top-six countries (by total tweets in our data). Consistent to the worldwide sentiment trends, the negative sentiment of all six countries dominates throughout. While a few countries (US, UK and India) reach a couple million negative tweets for a few weeks, the other countries stay lower around half a million in the remaining weeks. 

\begin{figure}[!h]
\centering    
\subfigure[United States positive sentiment]{\label{fig:us_pos_sent_box}\includegraphics[width=0.30\columnwidth]{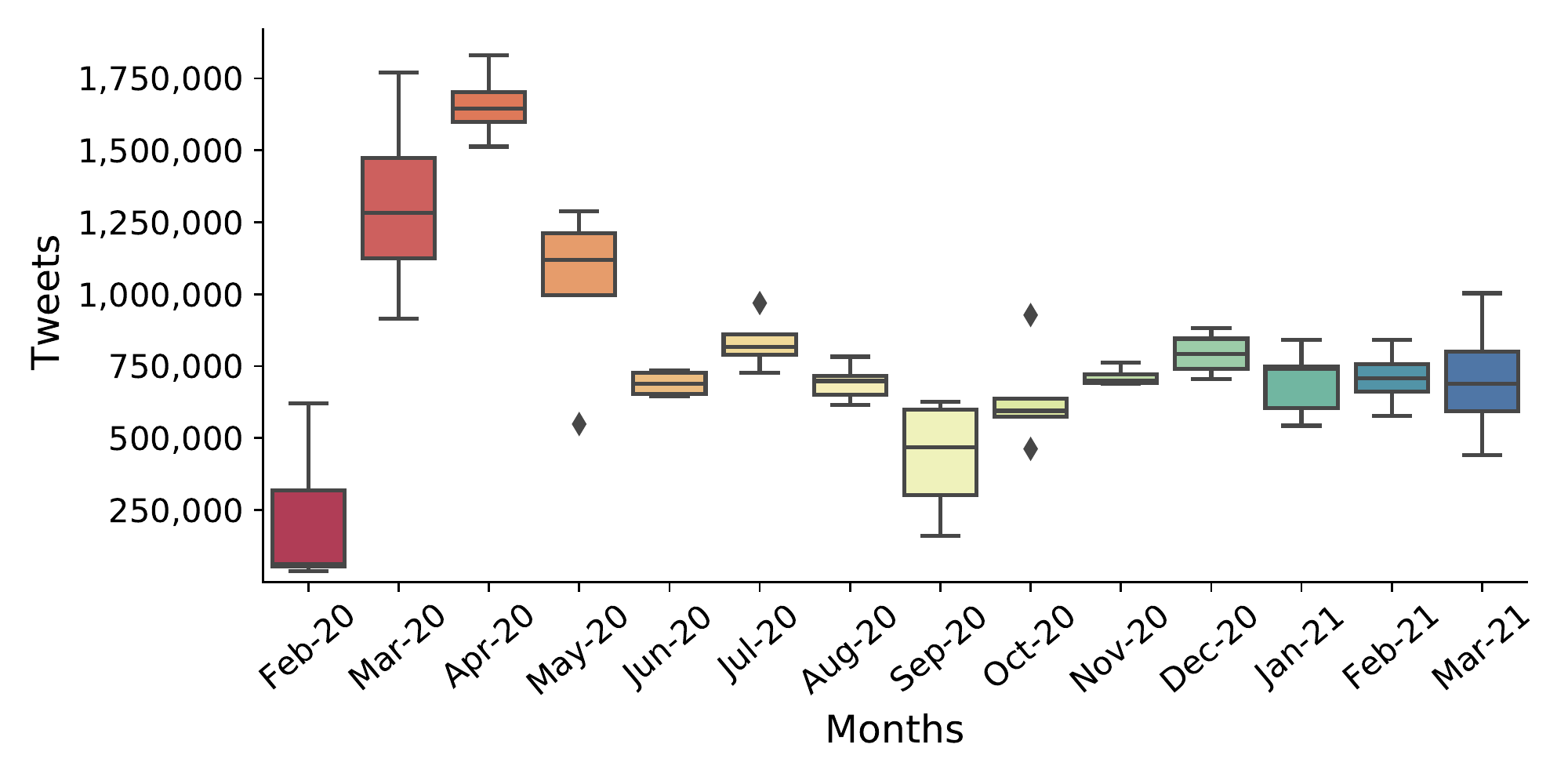}}
\subfigure[United States negative sentiment]{\label{fig:us_neg_sent_box}\includegraphics[width=0.30\columnwidth]{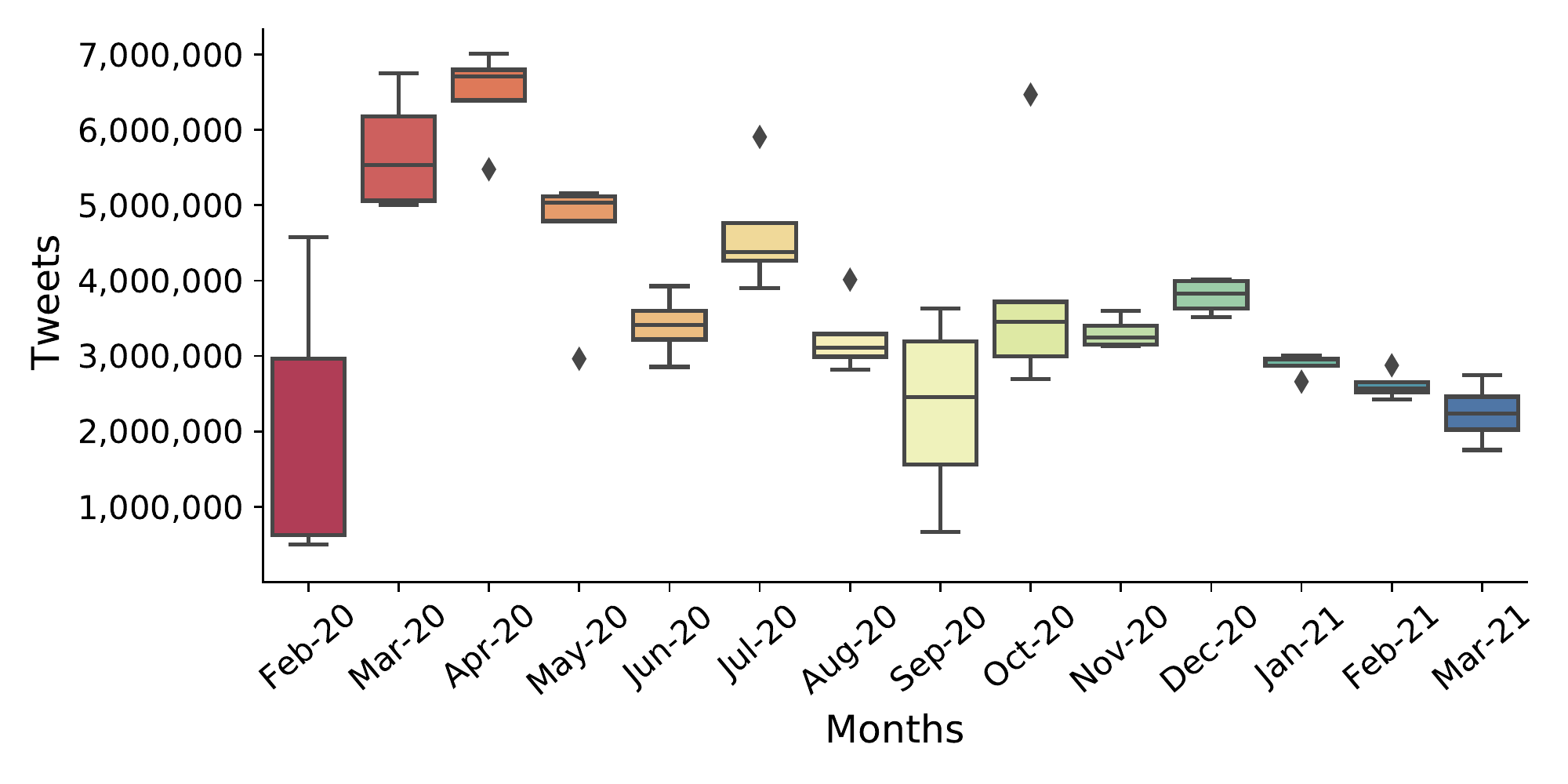}}
\subfigure[United Kingdom positive sentiment]{\label{fig:uk_pos_sent_box}\includegraphics[width=0.30\columnwidth]{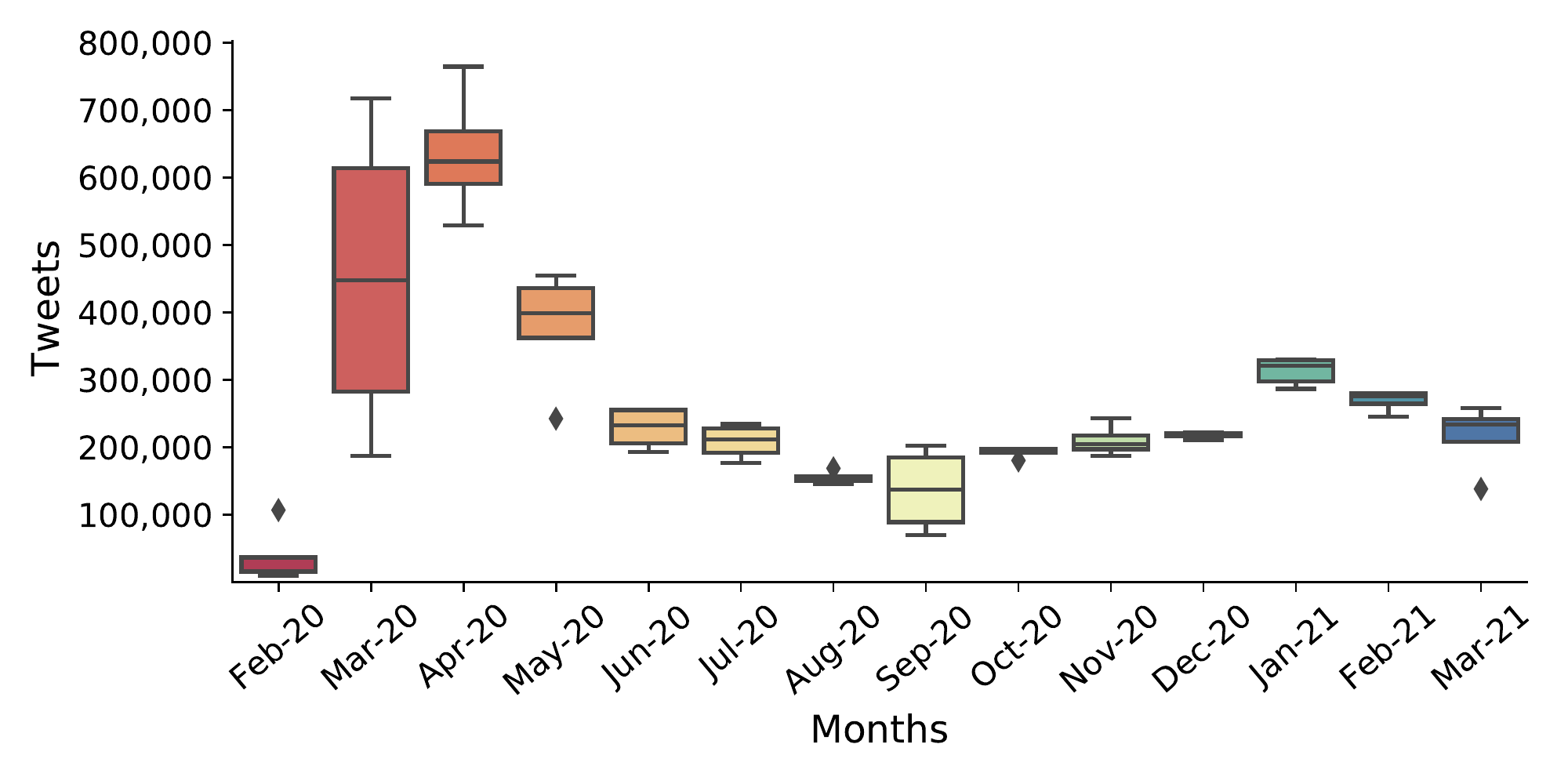}}
\subfigure[United Kingdom negative sentiment]{\label{fig:uk_neg_sent_box}\includegraphics[width=0.30\columnwidth]{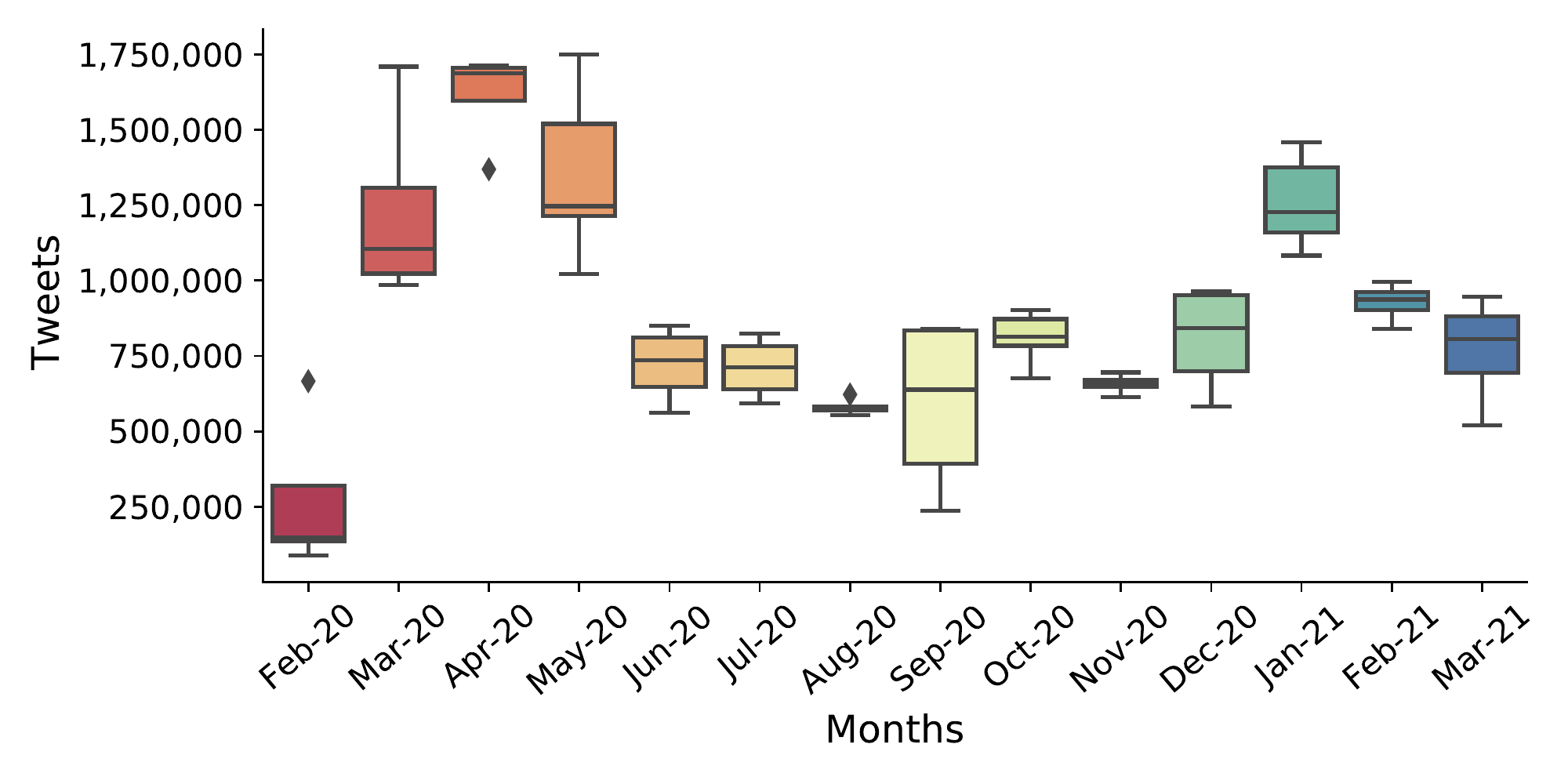}}
\subfigure[India positive sentiment]{\label{fig:india_pos_sent_box}\includegraphics[width=0.30\columnwidth]{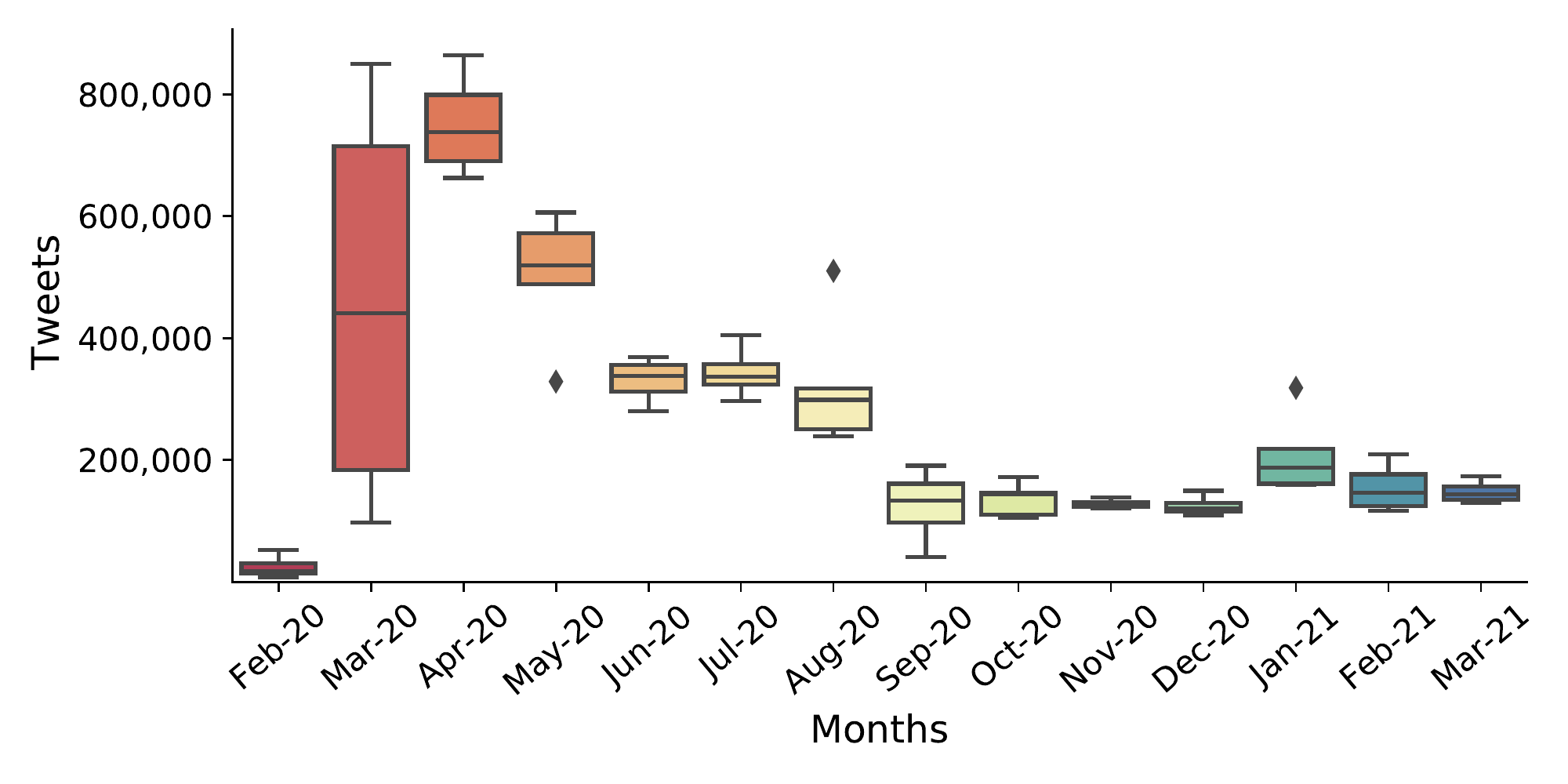}}
\subfigure[India negative sentiment]{\label{fig:india_neg_sent_box}\includegraphics[width=0.30\columnwidth]{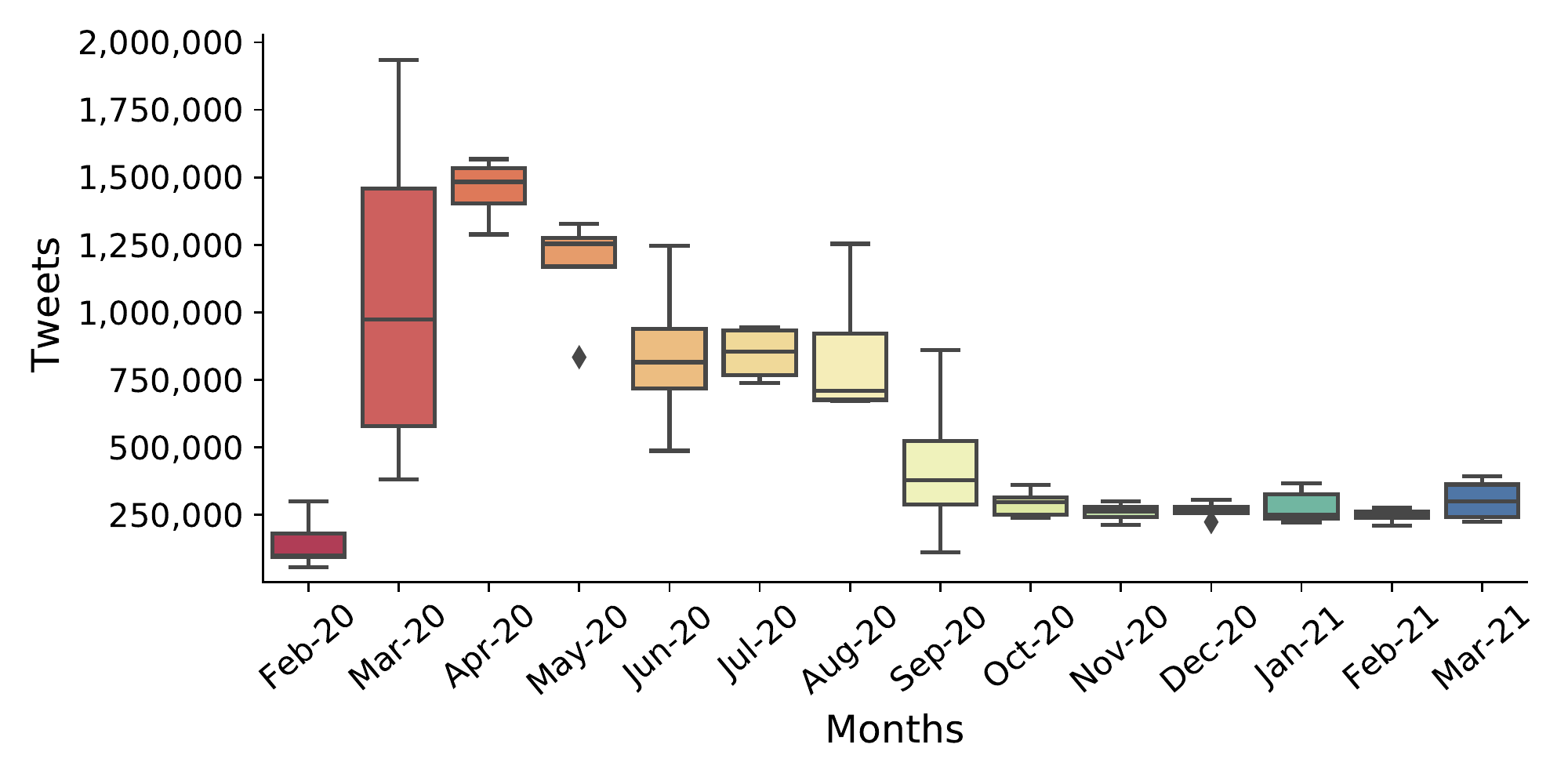}}
\subfigure[Canada positive sentiment]{\label{fig:canada_pos_sent_box}\includegraphics[width=0.30\columnwidth]{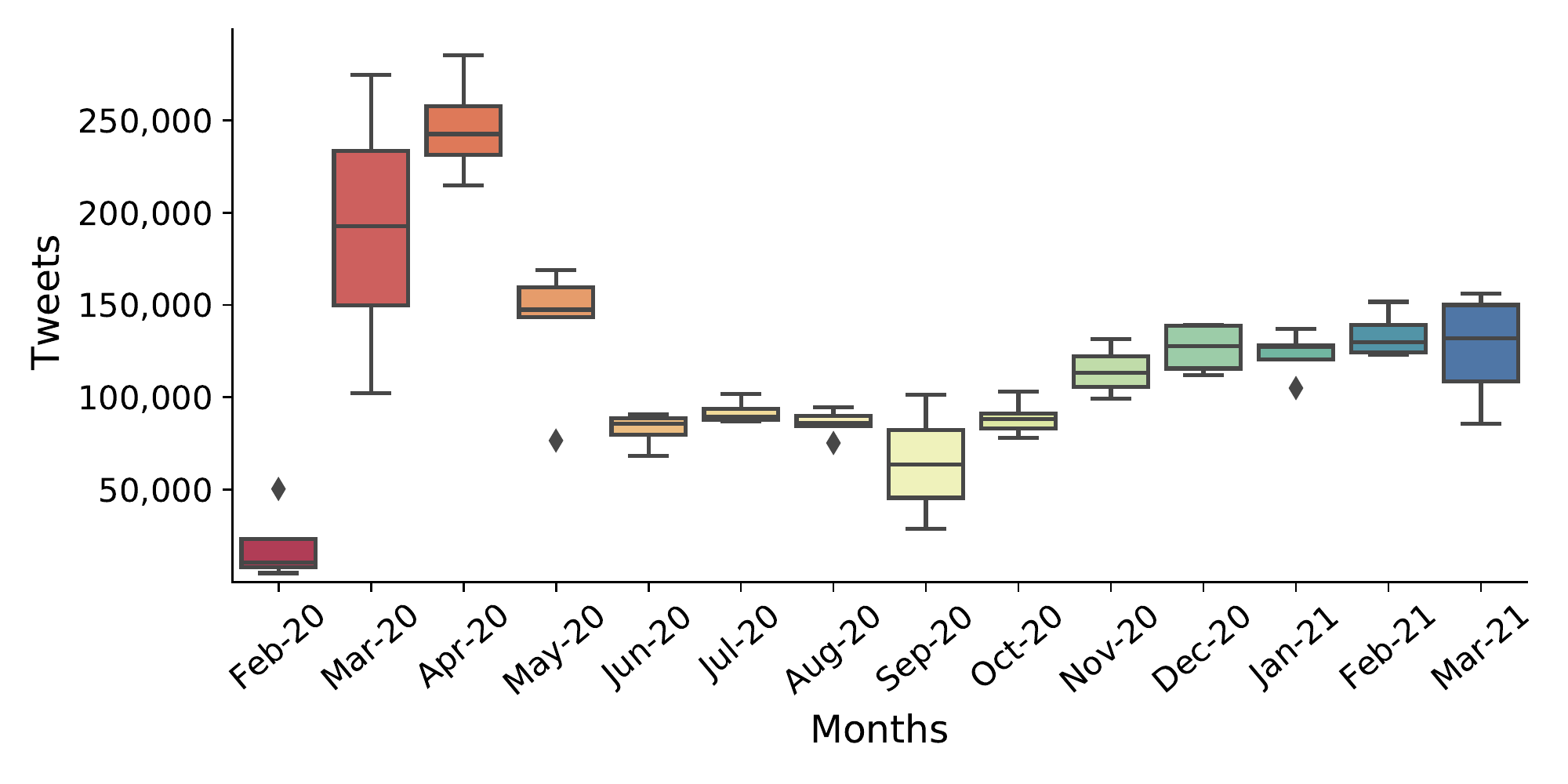}}
\subfigure[Canada negative sentiment]{\label{fig:canada_neg_sent_box}\includegraphics[width=0.30\columnwidth]{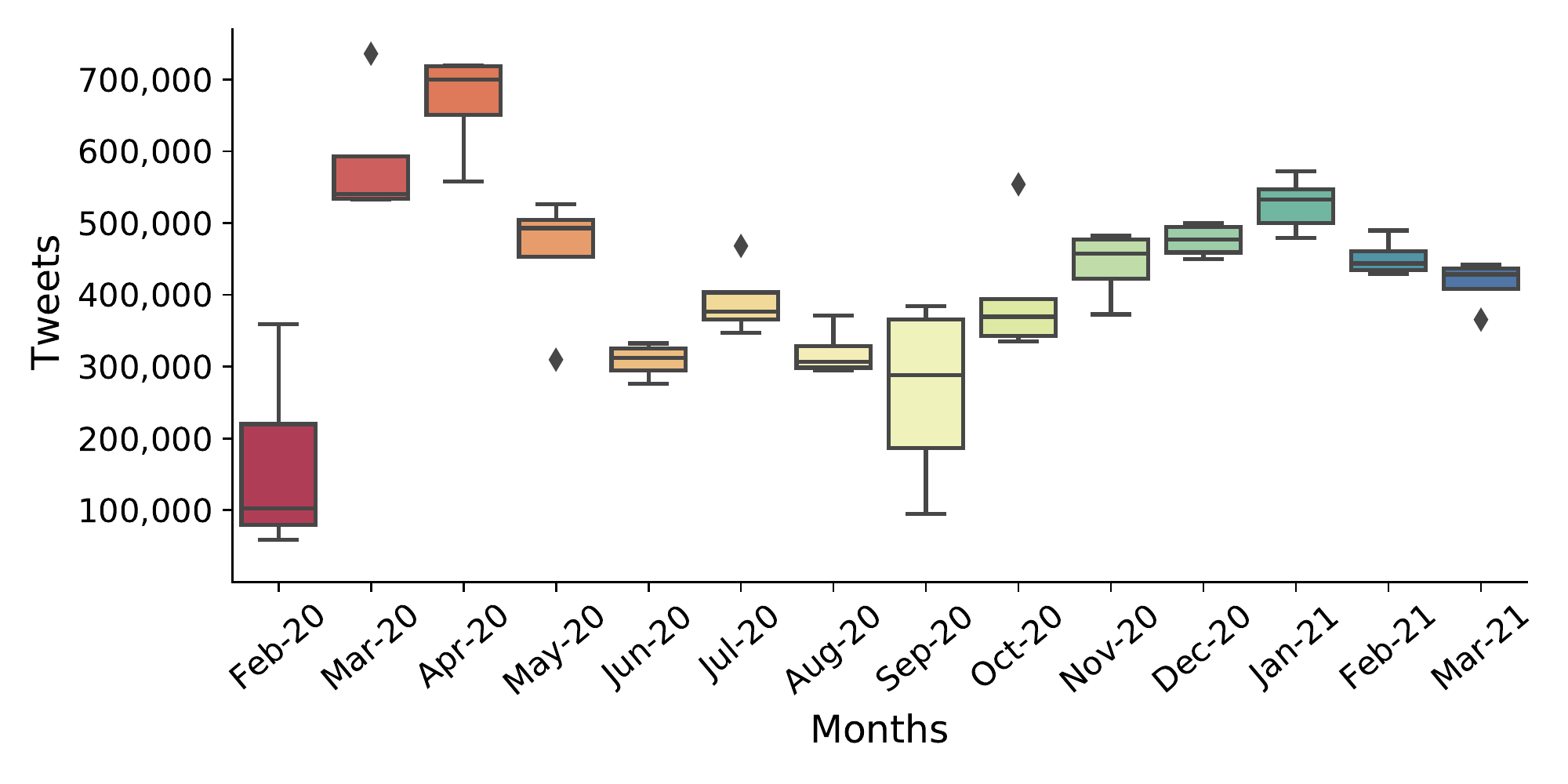}}
\subfigure[Brazil positive sentiment]{\label{fig:brazil_pos_sent_box}\includegraphics[width=0.30\columnwidth]{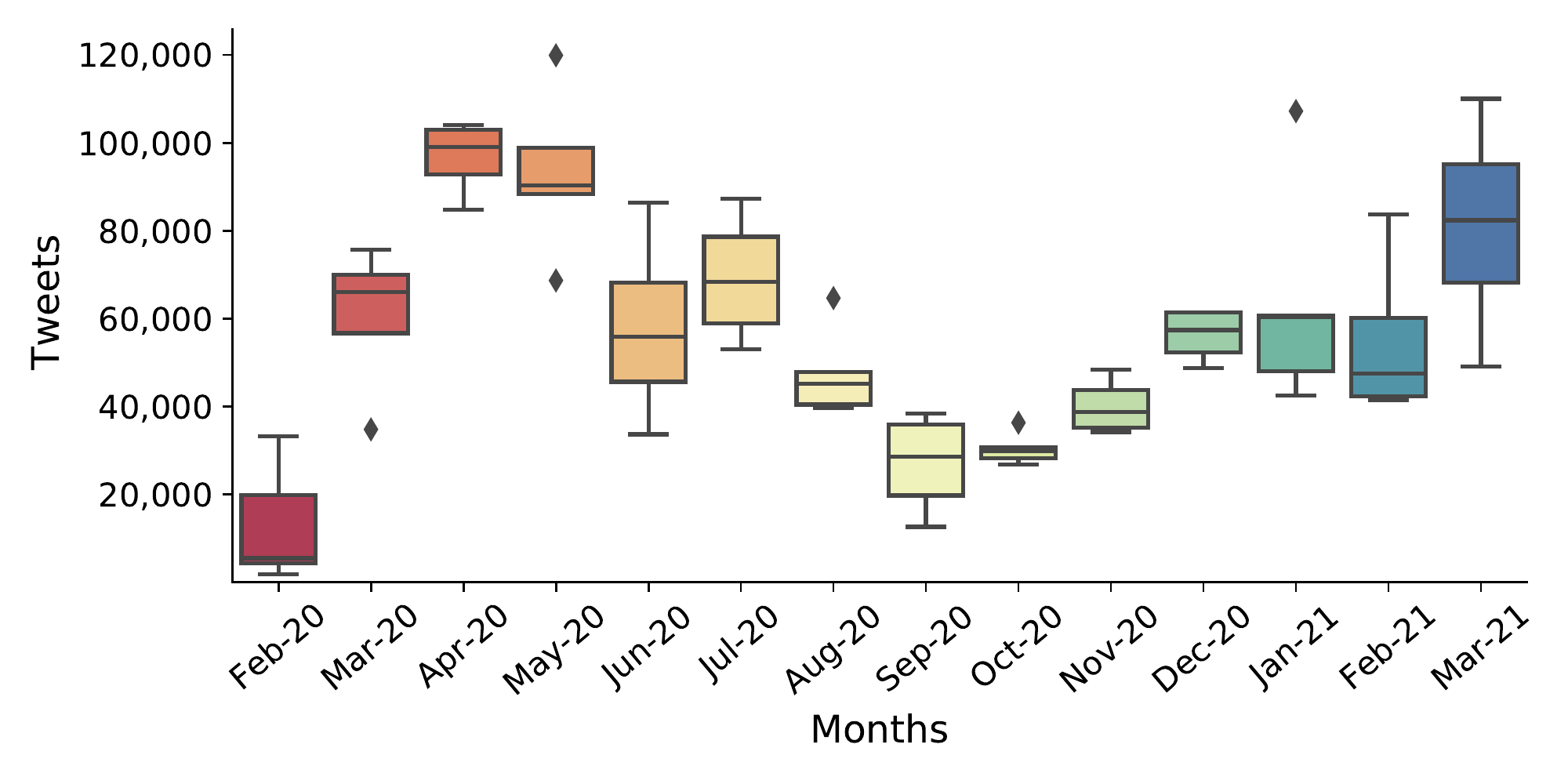}}
\subfigure[Brazil negative sentiment]{\label{fig:brazil_neg_sent_box}\includegraphics[width=0.30\columnwidth]{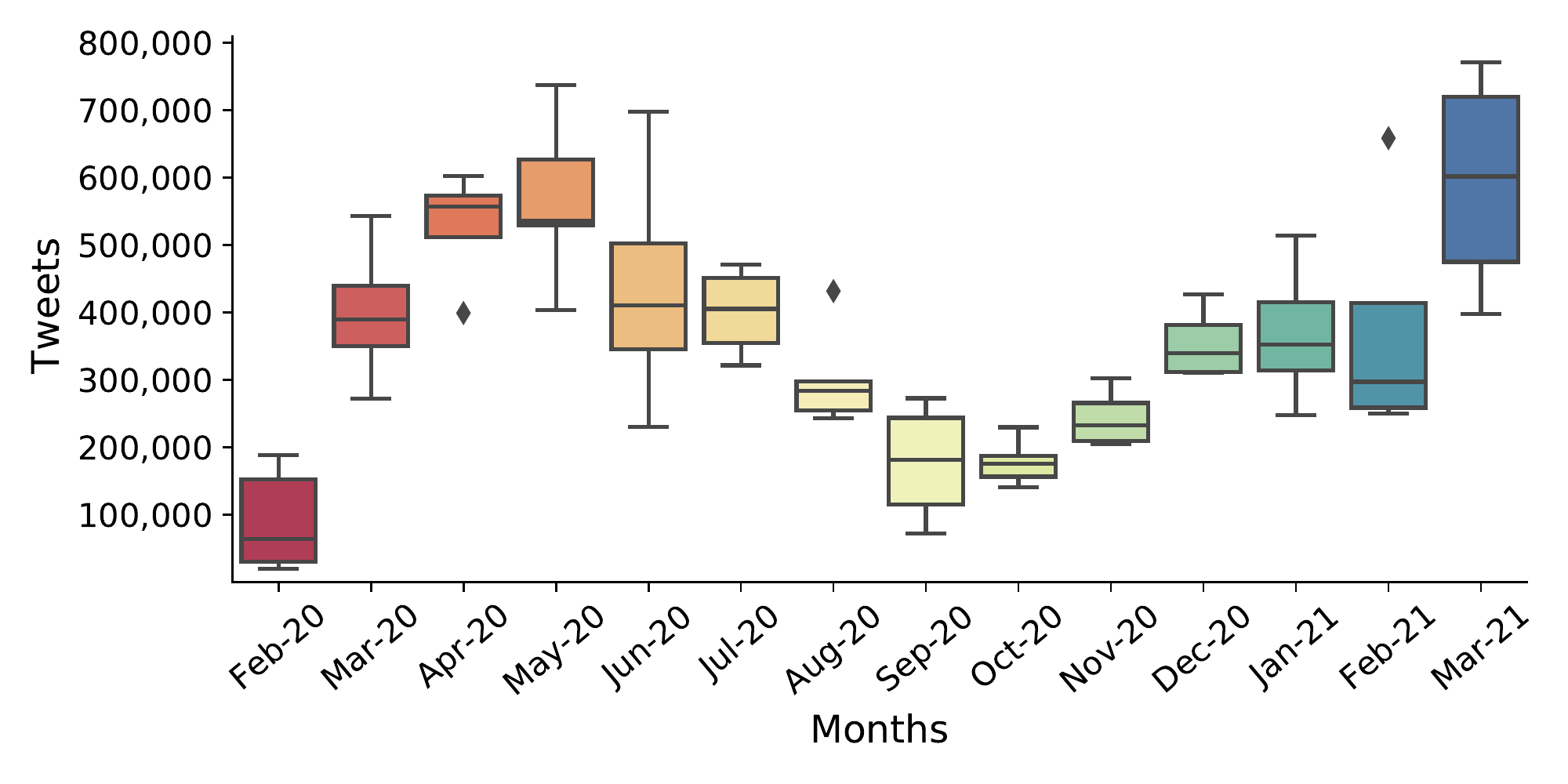}}
\caption{Monthly distribution summary of positive and negative sentiment tweets for the top five countries}
\label{fig:countries_sent_box}
\end{figure}

In Figure~\ref{fig:countries_sent_box}, we provide additional information about the distribution, skewness through quartiles, and median for positive and negative sentiments for the top-five countries. We notice that in most cases the earlier months of COVID-19 (i.e., February-March 2020) show high variations in both positive and negative sentiments, except for UK and India, where the number of both positive and negative sentiments are comparatively low with high level of agreement with each other. Surprisingly, the February 2020 data for both types of sentiments in the US and especially negative sentiment in other countries is highly positively skewed. Most countries seem to have less dispersion in April 2020 with quite high maximum range of any type of sentiment. These interesting patterns can reveal many more hidden insights, which could help authorities gain situational awareness leading to timely planning and actions. 

\begin{figure}[!ht]
\centering
\includegraphics[width=0.85\textwidth]{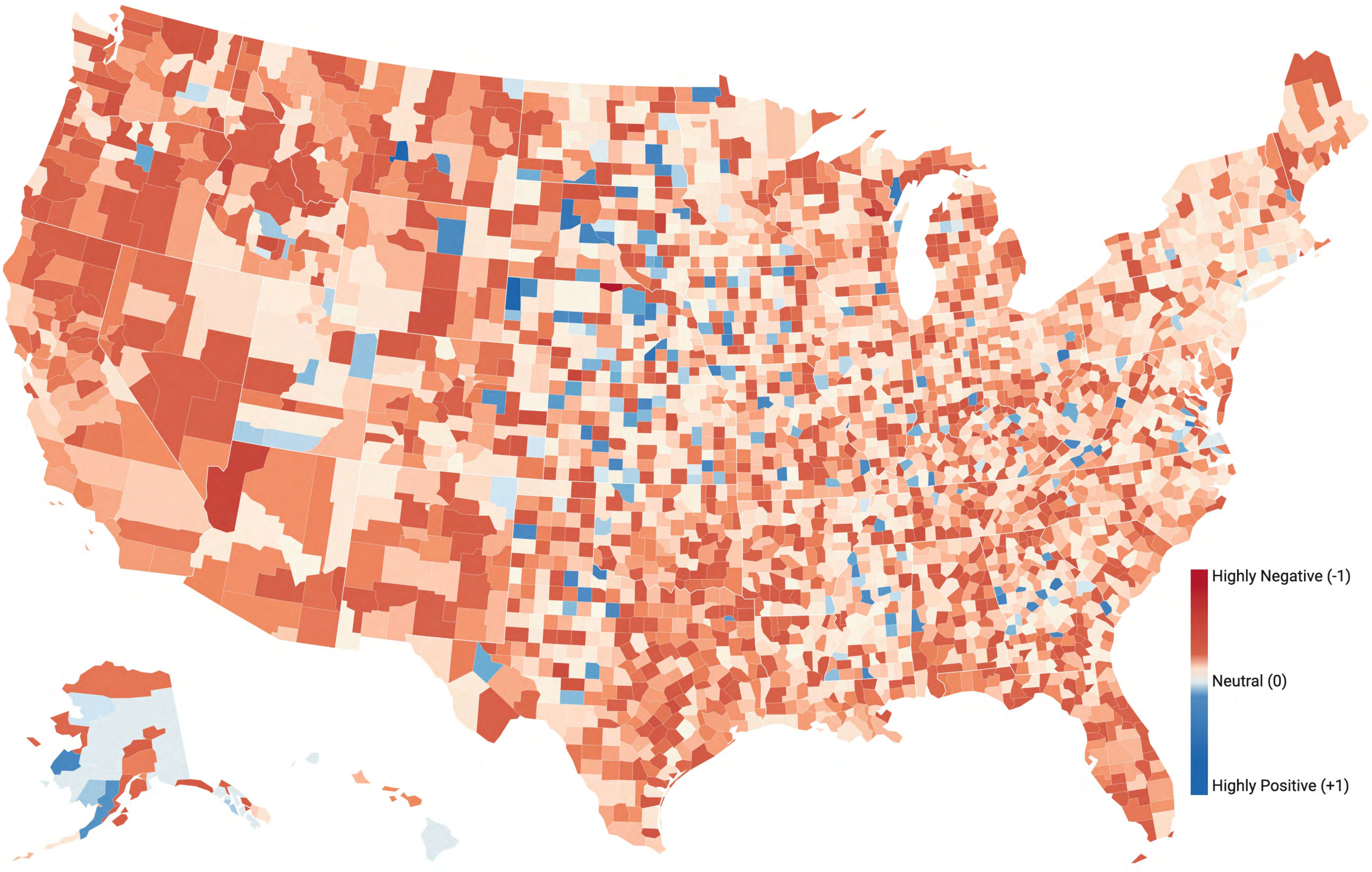}\vspace{-0.4cm}
\caption{Sentiment across US counties. Tweets geotagged using \textit{user location}, \textit{user profile description}, and \textit{GPS-coordinates} are used after normalizing by the total number tweets from each county.}
\label{fig:us_county_sent_map}
\end{figure}

Figure~\ref{fig:us_county_sent_map} shows the distributions of sentiment scores across the US counties. Similar to the worldwide sentiment map, the sentiment scores for counties are normalized by the total number of tweets from each county using the weighted average for positive, negative, and neutral tweets. Overall, the negative sentiment dominates across different states and counties. While most counties show strong to moderate negative sentiment, a strong positive sentiment can be observed for the Sioux County in Nebraska, Ziebach County in South Dakota, Highland County in West Virginia, and Golden Valley County in Montana. California is mostly on the negative side whereas New York appears near neutral or on the negative side. Texas seems to represent all ends of the spectrum---covering moderate-to-strong negative as well as some positive sentiment. Florida and Washington are all negative. Overall, the western region is mostly negative, the Midwest is fairly divided but strong in whatever sentiment it exhibits, the Northeast region shows less negative intensity (more towards neutral), and the Southern region shows some counties with positive sentiment, but the majority is either negative or neutral. 

\begin{figure}[!ht]
\centering
\includegraphics[width=0.80\columnwidth]{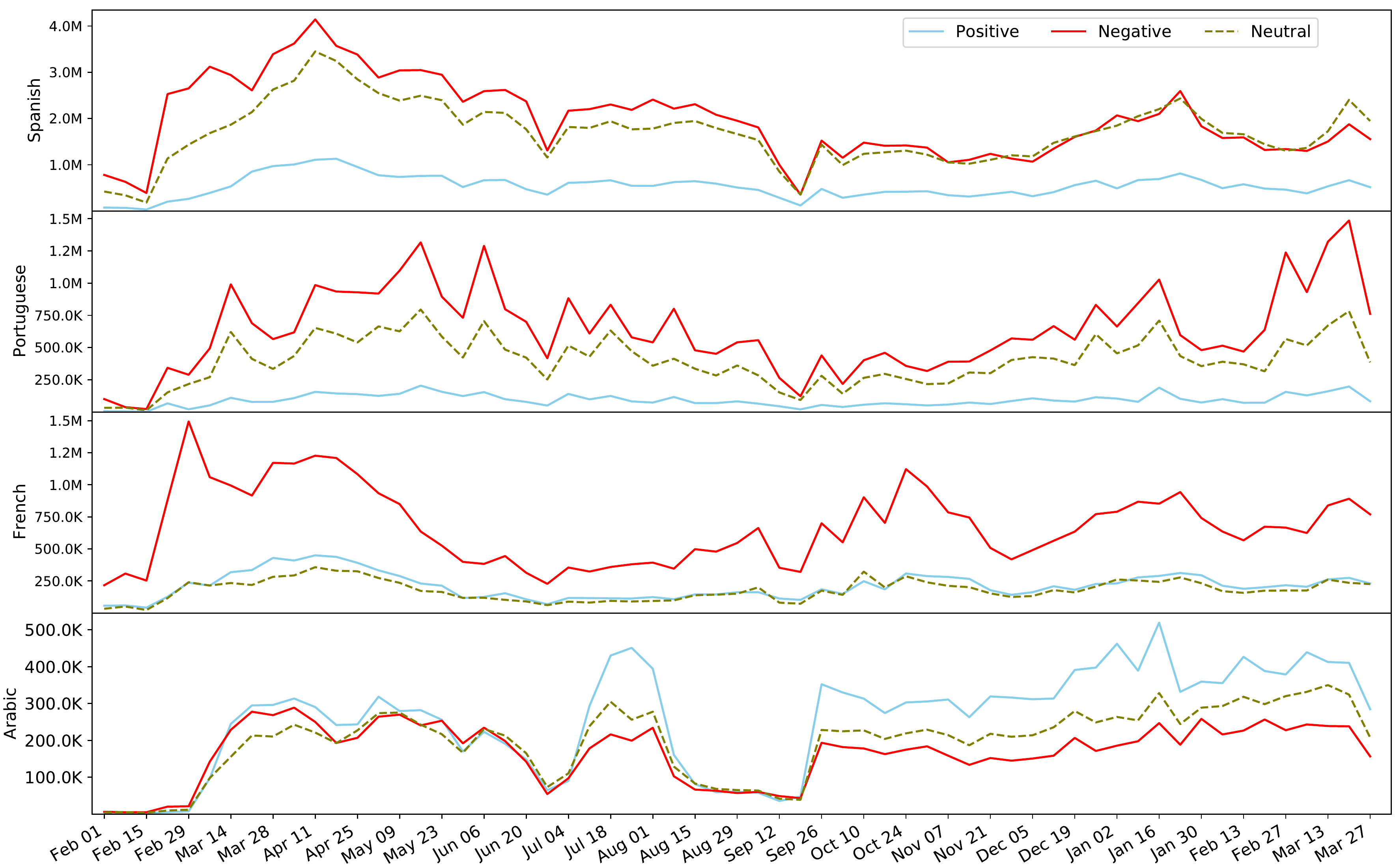}
\caption{Weekly distribution of sentiment labels of tweets in four languages (Spanish, Portuguese, French, and Arabic).}
\label{fig:langs_sentiment}
\end{figure}

Figure~\ref{fig:langs_sentiment} presents the distributions of sentiment labels for four languages. Interestingly, the Arabic language shows the domination of the positive sentiment throughout the 14 months except February 2020 and a few weeks in the middle. For the other three languages, the negative sentiment surpasses the other two sentiment classes. While all show peaks in and around April and May 2020, the surge of the negative sentiment in February and March 2021 in the case of Portuguese is noticeable and requires further investigation. 

\subsection*{User type and gender classification}
Twitter has 186 million daily active users with 70.4\% male and 29.7\% female users~\cite{twitter_stats}. Twitter users represent, among others, businesses, government agencies, NGOs, bots, and---most importantly---the general public~\cite{zhang2019less,uddin2014understanding}. Information about user types is helpful for many application areas, including customer segmentation and engagement~\cite{okazaki2015using}, making recommendations~\cite{hannon2010recommending}, users profiling for content filtering~\cite{garcia2013catstream}, and more. Moreover, users demographic information such as gender is important for addressing societal challenges such as identifying knowledge gaps~\cite{manierre2015gaps}, health inequities~\cite{johnson2009better}, digital divide~\cite{antonio2014gender}, and other health-related issues~\cite{lawrence2007methodologic}. The tweets in TBCOV are from 87.7 million unique users worldwide, which is 47\% of the daily active users on Twitter. Our aim is to determine accounts which belong to the general public, hereinafter \textit{personal accounts}, and their gender. However, Twitter neither provides account types nor their gender information. To this end, we observed that user-provided names in \textit{personal accounts} can potentially be used to not only distinguish them from other types such as \textit{organizational accounts}, their morphological pattern are indicative of gender as well~\cite{ali2019morphological,slepian2016voiced}. For example, the username ``Capital Press'' is a media account whereas the username ``Laura Sanchez'' is a personal account that likely belongs to a female.

First, we determine users' type (i.e., \textit{personal, organizations, etc.}) by applying the English NER model (described previously) on user-provided names. Usernames are preprocessed (i.e., remove URLs, numerals, emojis, tabs spaces, newlines) prior to feeding the model, which assigns one of the eighteen entity types to a username, including \textit{person}. Entity types of all 87.7 million usernames are obtained according to which there are 46,504,838 (52.98\%) \textit{person}, 11,909,855 (13.57\%) \textit{organization}, and 29,357,141 (33.45\%) \textit{miscellaneous} user types. More importantly, nearly half (48\%) of the tweets in the dataset are posted by \textit{personal accounts}, 11\% by organizational, and 40\% by other user types.  

\begin{figure}[!h]
\centering
\includegraphics[width=\columnwidth]{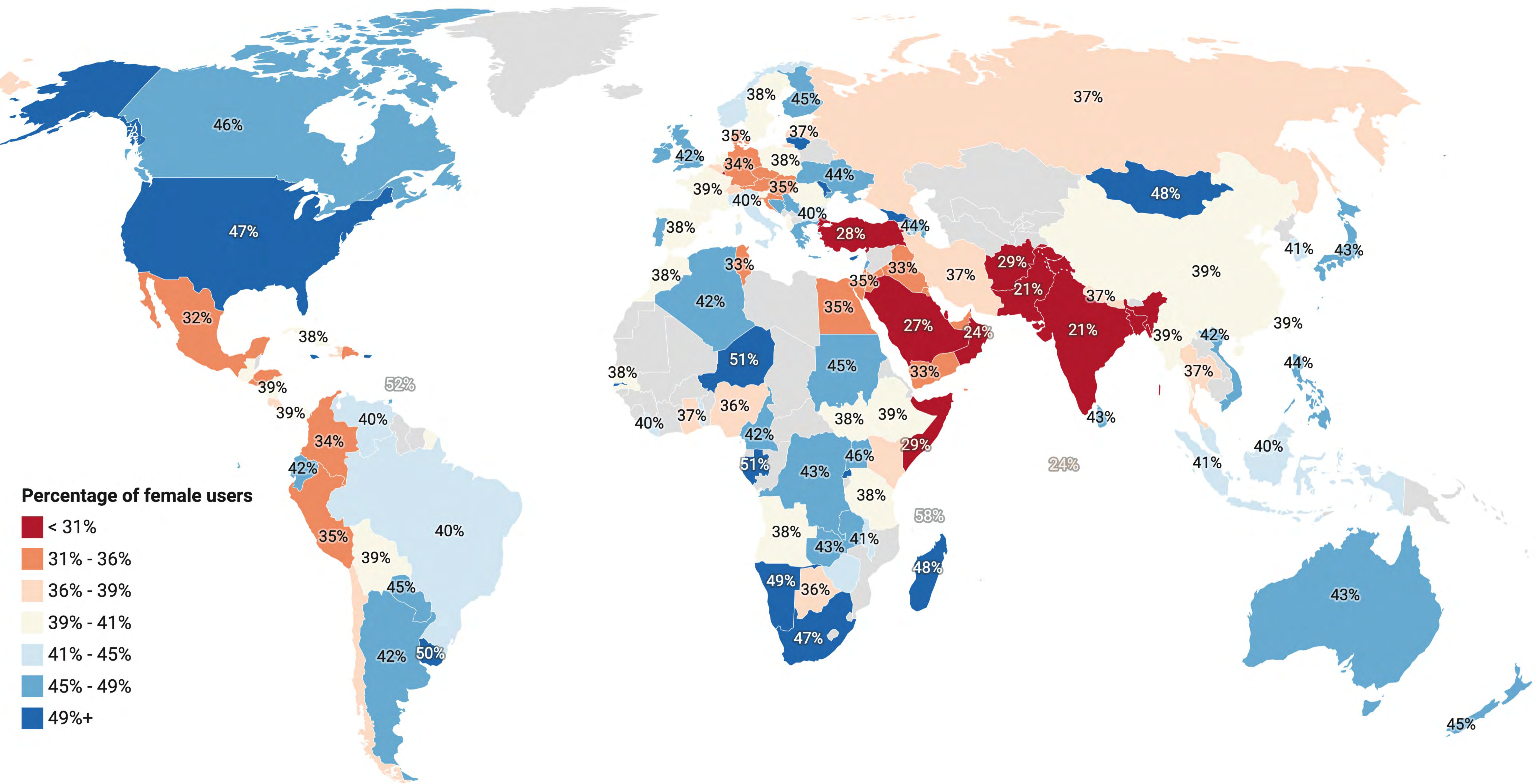}
\caption{Percentage of female users for countries meeting representative sampling criteria (confidence interval=95\%; margin of error$\leq1\%$). Gray color indicates the countries excluded due to under representation ($N=85$).}
\label{fig:female_users}
\end{figure}

Next, we sought to further disaggregate the identified \textit{personal accounts} (i.e., 46,504,838) by their gender. Prior studies demonstrate that morphological features of a person's \textit{given} name (also known as a \textit{first} name or \textit{forename}) provide gender cues, such as voiced phonemes are associated with male names and unvoiced phonemes are associated with female names~\cite{slepian2016voiced}. Hence, the first names of the identified \textit{personal accounts} are employed for training supervised machine learning classifiers. Several publicly available name-gender resources were used~\cite{dataw_gender_data, cmu_gender_data, dataw_gender_data2} as our training datasets. Names in these datasets are written using the English alphabets. We combined the datasets and removed duplicates. This process yielded 121,335 unique names with a distribution of female and male as 73,314 (60\%) and 48,021 (40\%), respectively. 

Prior to training classifiers, data was split into train and test sets with a 80:20 ratio, respectively, and phonetic features from \textit{first names} are extracted by moving a variable-sized window over them in two directions (i.e., left-to-right and the opposite). The window of length one moves from its starting point (i.e., either the first or the last character of a name). Subsequent moves increases window size by one until a threshold value reached. The threshold limits the number of features required in one direction, which we empirically learned by experimenting several values ranging from 1-to-7 (i.e., 7 is the average length of names in our dataset). Fewer than four features (in one direction) negatively impact classifiers' performance, whereas, larger values yield diminishing effect. Thus, a threshold of four is set, i.e., representing the first four and last four features of a name.  For example, given a name \textit{``Michael''}, the feature extraction method extracts eight features, four from the start (i.e., \textit{`m', `mi', `mic', `mich'}) and four from the last (i.e., \textit{`l', `el', `ael', `hael'}). The extracted features are then encoded with their corresponding positions in names, e.g., the \textit{`mic'} feature in the earlier example caries its position i.e., \textit{first-three-letters}. The extracted positional features are then used to train several well-known machine learning classifiers, including Naive Bayes\cite{rish2001empirical}, Decision Trees\cite{quinlan1986induction}, and Random Forests\cite{breiman2001random}. The Random Forests algorithm yields better performance, and thus, used to process all 87.7 million names. The evaluation of gender classification model is presented in the next section. 

The gender classification process identified 19,598,252 (72.84\%) female and 26,906,586 (57.86\%) male users. Although the proportion of female users is higher than the male users, the number of tweets posted by the male users is 15\% more than the female users. Specifically, of all 963,681,513 tweets from \textit{personal accounts}, 558,259,178 (57.93\%) are from male and 405,422,335 (42.07\%) from female users. We further determine female to male ratios for each country. To choose countries for computing female to male ratios, we estimated the required sample size for each country. We set our confidence interval at 95\% and margin of error to $\leq$1\%. Countries with users (any gender) less than the required sample size are dropped ($N=78$). Figure~\ref{fig:female_users} shows the percentage of female users for countries meeting the representativeness criteria.

\begin{figure}[!h]
\centering
\includegraphics[width=\textwidth]{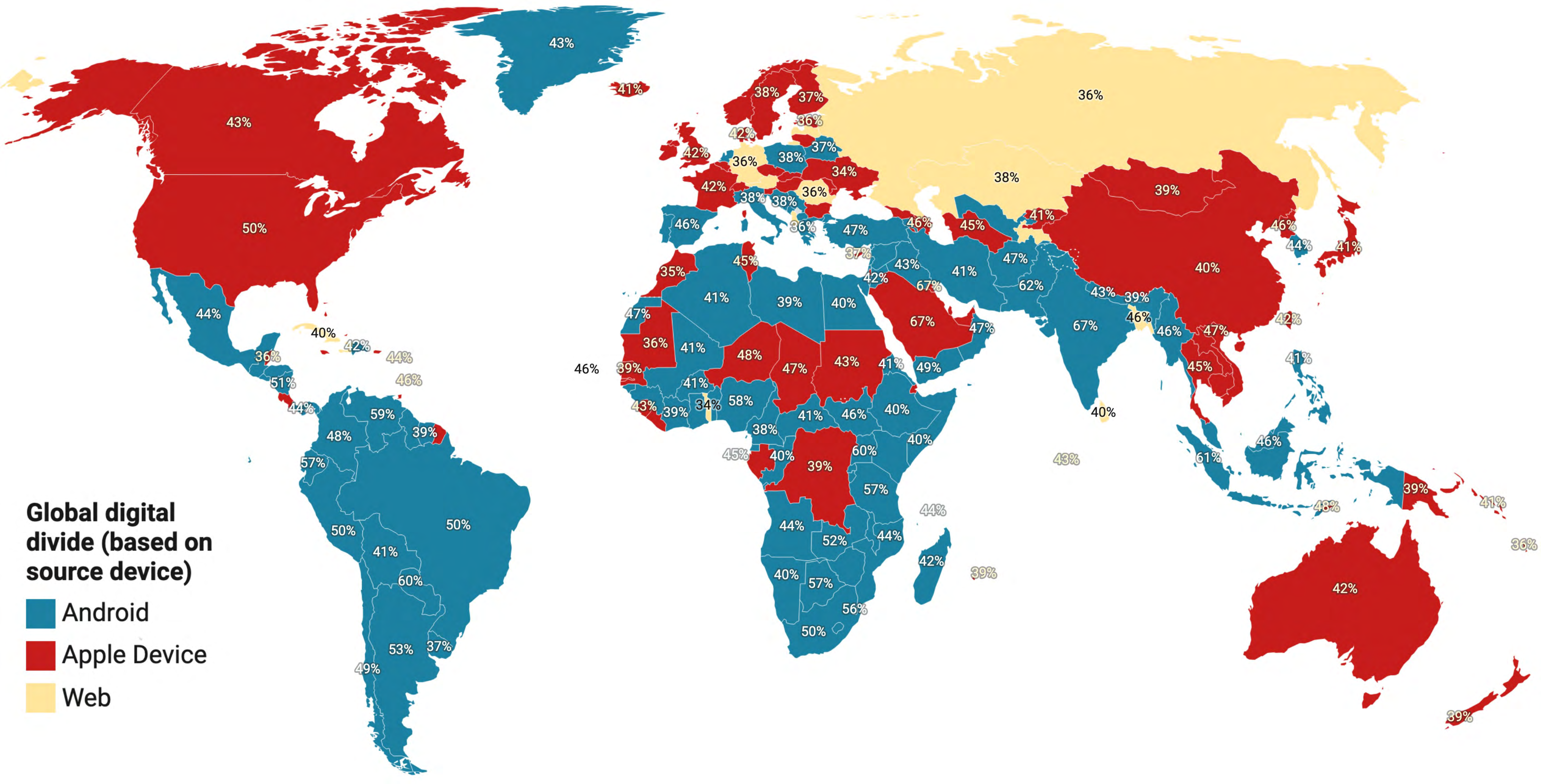}
\caption{Global digital divide estimated through the type of device used for tweeting. Representative device type penetration (percentage) is shown on top of each country.}
\label{fig:digital_divide_map}
\end{figure}

\subsubsection*{Global digital divide }
Next, we sought to determine global digital divide by relying on users access to different types of devices used for tweeting. Out of all more than two billion tweets, we extracted 1,003 unique application types (provided by Twitter) supporting the tweet posting feature. Dozens of applications support tweeting feature, including both web-, and mobile-based apps. 
We manually analyzed all the applications to determine the operating system they are built for (e.g., iOS, Android). Next, based on the operation system information, we categorized each application into one of the three device types i.e., \textit{(i) Apple device}---representing all iOS devices such as iPhone, iPad, etc., \textit{(ii) Android}---representing all types of Android-based devices, and \textit{(iii) Web}---representing all the web-based applications for tweeting. Finally, an aggregation is performed on device types for each county and the most frequent device is selected. 

Figure~\ref{fig:digital_divide_map} shows the most frequently used device type in each country. The map shows a device type for 217 countries worldwide. Of all, the Android is the most used device type with $N=103$ (48\%), Apple with $N=97$ (45\%), and Web is the least used with $N=17$ (7\%). As Apple devices are more expensive than Android, we expect to see Apple's domination in rich countries. This assumption stands true except a couple of countries, including Niger and Senegal, among others.  

\subsection*{Trends Analysis}
\label{sec:trends}
The impact of the COVID-19 pandemic on people's livelihoods, health, families, businesses, and employment is devastating. To determine whether \textit{TBCOV} covers information about such unprecedented challenges, next we perform trend analysis of six important issues. The first two issues are directly related to people's health, i.e., (i) tweets about anxiety and depression, and (ii) self-declared COVID-19 symptoms. Next two issues represent severe consequences of COVID-19 that millions of families worldwide directly faced, i.e., (iii) deaths of family members and relatives, and (iv) food shortages. The last two issues are about people's social life and preventive measures, i.e., (v) face mask usage in public areas as well as shortages, and (vi) willingness to take or already taken vaccine. 

For each issue, a set of related terms are curated to form logical expressions. For instance, in the case of the ``COVID19 symptoms'' issue, we divide it into five sub-groups representing different COVID-19 symptoms listed on the CDC website\footnote{https://www.cdc.gov/coronavirus/2019-ncov/symptoms-testing/symptoms.html}, which can also be seen below in Table~\ref{tab:trend_groups}. Several related terms were added to each sub-group to increase the recall. For example, for COVID deaths of parents, the ``parents'' group contains two sets of terms: (i) ``father OR mother OR dad OR mom'', and (ii) ``deceased OR succumbed OR perished OR lost battle OR killed OR my * passed OR my * died''\footnote{Asterisk (*) allows one term from set (i) to appear in between}. The logical operator `AND' between these two sets forms the final expression used to retrieve weekly tweets. The full list of terms will be released with the dataset.

\begin{table}[!h]
\centering

\footnotesize
\begin{tabular}{ll}
\toprule
\textbf{Topics representing different issues} & \textbf{Sub-topics related to the main topic} \\
\midrule
COVID-19 symptoms & Fever, cough, shortness of breath, headache, loss of taste and smell \\
COVID deaths mentions & Parents, siblings, grandparents, relatives, and close connections \\
Food shortages & Food availability, food access, food adequacy, and food acceptability \\
Anxiety \& depression & Anger, sleepless, fearful, upset, restless, and anxious\\
Mask usage \& importance & Mask violation, masks are important, wear masks, masks save lives, masks useless\\
Willingness to take/taken vaccine & Reactions to vaccine, harmful vaccine, got vaccine, covid jab taken, will take vaccine\\
\bottomrule
\end{tabular}
\caption{Term groups of four topics for trend analysis}
\label{tab:trend_groups}
\end{table}

\begin{figure}[!h]
\centering    
\subfigure[COVID-19 symptoms]{\label{fig:symptoms_trends}\includegraphics[width=0.49\columnwidth]{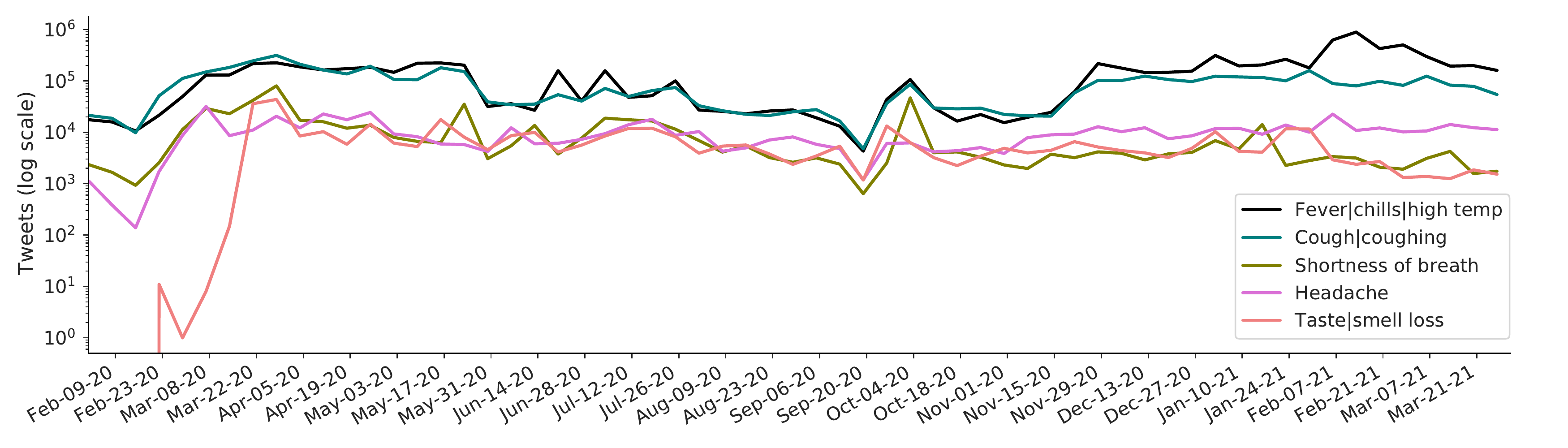}}
\subfigure[Anxiety, anger, sadness]{\label{fig:axiety_trends}\includegraphics[width=0.49\columnwidth]{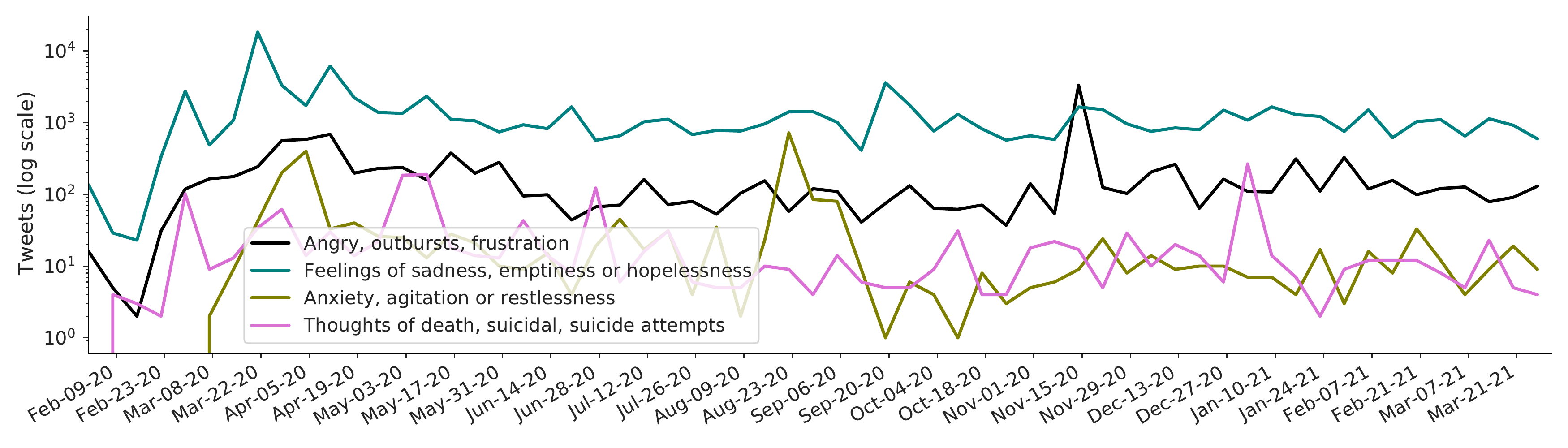}}
\subfigure[Parents, siblings, relatives deaths]{\label{fig:deaths_trends}\includegraphics[width=0.49\columnwidth]{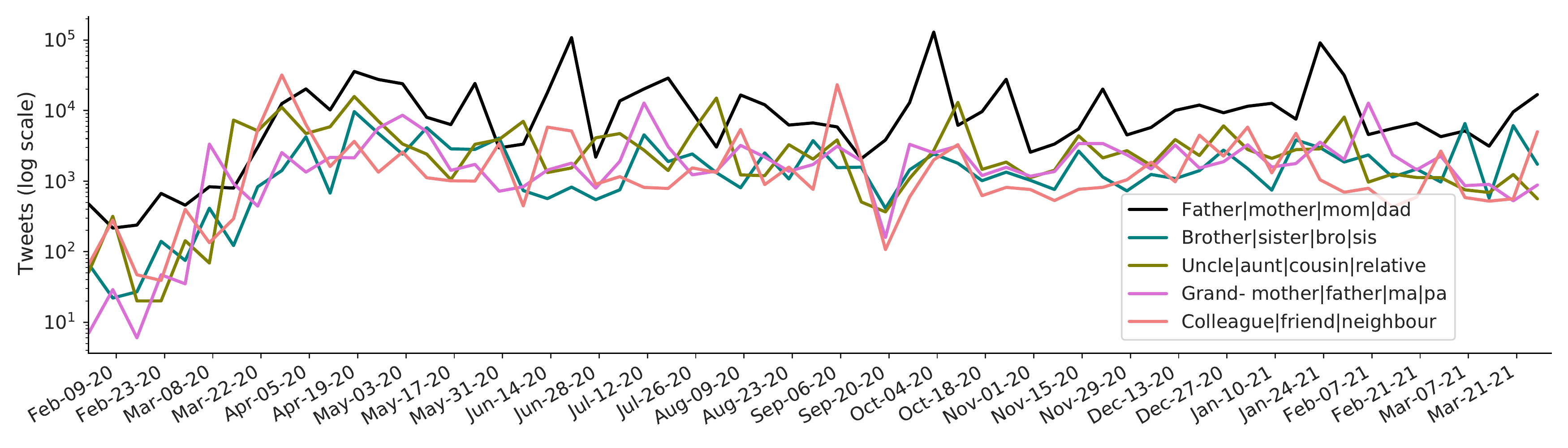}}
\subfigure[Food insecurity and shortages]{\label{fig:food_trends}\includegraphics[width=0.49\columnwidth]{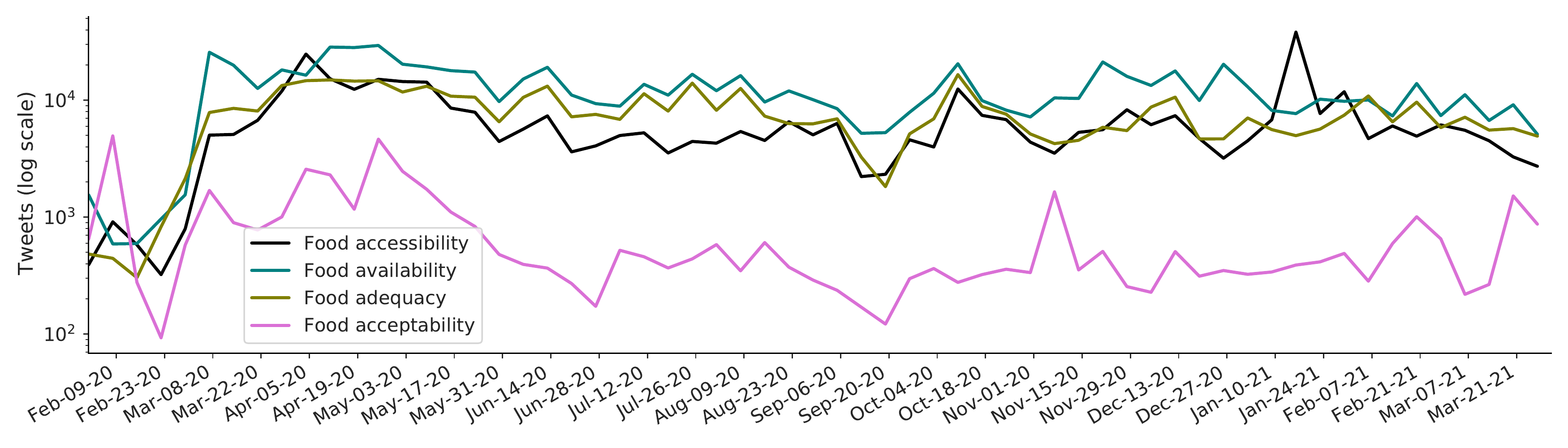}}
\subfigure[Mask usage \& shortages \& importance]{\label{fig:mask_trends}\includegraphics[width=0.49\columnwidth]{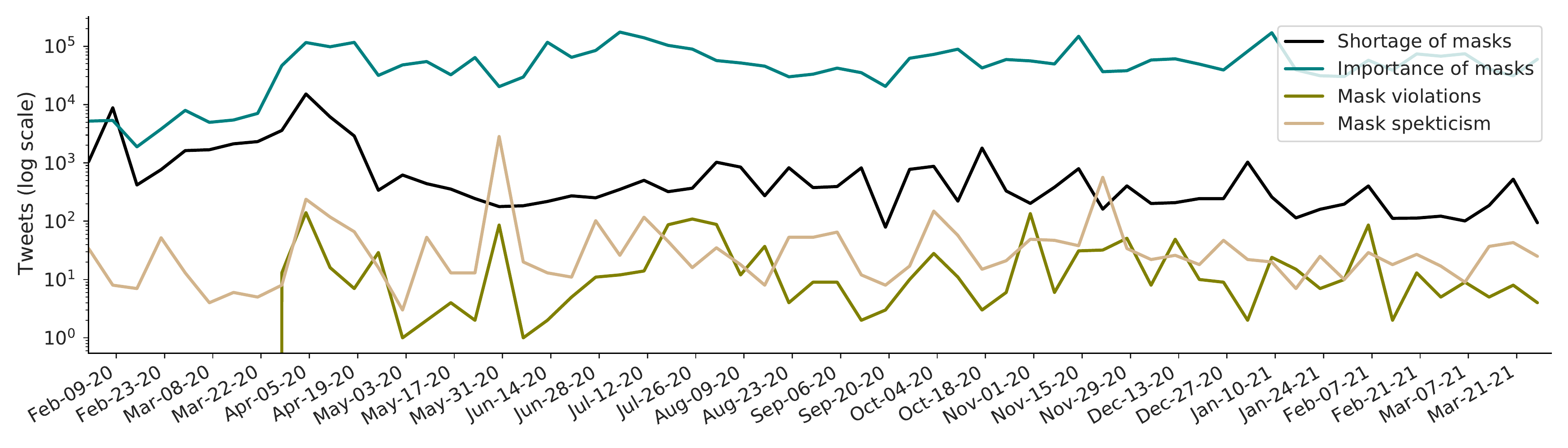}}
\subfigure[Willingness to take or taken vaccine]{\label{fig:vaccine_trends}\includegraphics[width=0.49\columnwidth]{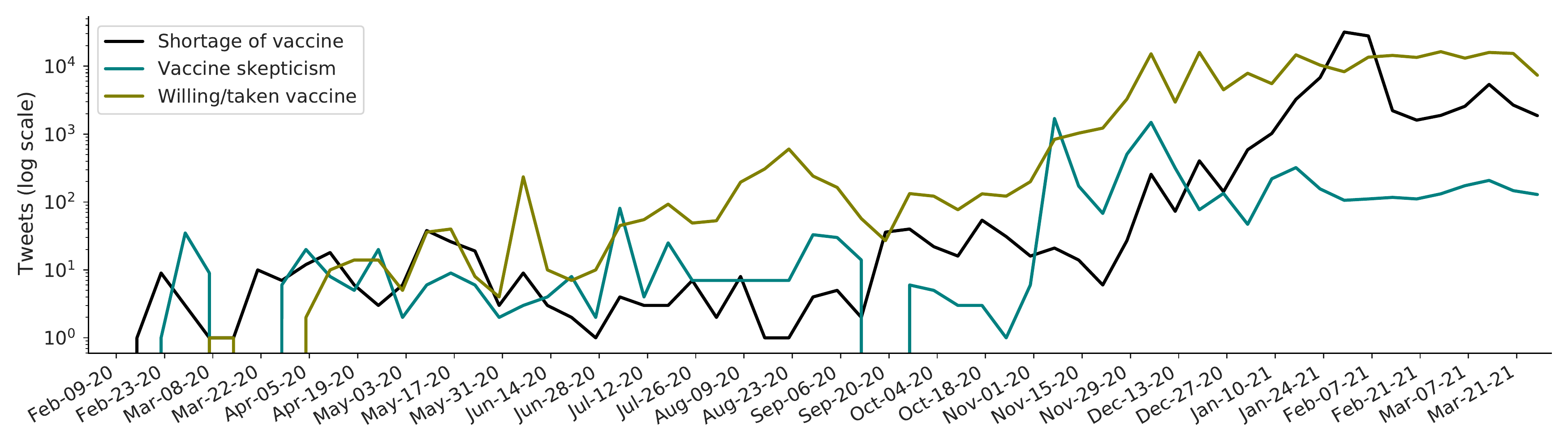}}\vspace{-0.4cm}
\caption{Weekly trends of important issues related to personal and social lives of users linked to COVID-19\vspace{-0.1cm}}
\label{fig:weekly_trends_misc_topics}
\end{figure}

Figure~\ref{fig:weekly_trends_misc_topics} depicts weekly distributions (in log scale) of the retrieved tweets. Figure~\ref{fig:symptoms_trends} shows sub-groups of the COVID-19 symptoms category. The two most reported symptoms in tweets are \textit{fever} and \textit{cough} followed by the \textit{shortness of breath} and \textit{headache}. Interestingly, reports of \textit{loss of taste} and \textit{smell} are almost zero until the end of February 2020, which then suddenly spike from March 8th onward. 
Figure~\ref{fig:axiety_trends} shows trends of different groups for the anxiety and depression topic. The feelings of sadness and hopelessness seem to dominate throughout the year followed by anger, outburst, and frustration. Surprisingly, the expressions with suicidal thoughts are captured in the data, as well. These particular trends need an in-depth investigation to better understand motives behind such extreme thoughts for authorities to intervene and offer counseling.

The weekly trends representing two important and direct consequences of COVID-19 on the general public are shown in Figure~\ref{fig:weekly_trends_misc_topics}(c \& d), i.e., tweets mentioning death of parents, siblings, relatives or close connections; and food insecurity in terms of its availability, accessibility, adequacy, and acceptability. A large number of tweets reporting deaths is observed with majority about parents. Grandparents and the category representing uncle and aunt are significant as well. Overall, elderly death reports are significantly higher than younger population. 

Similarly, TBCOV shows coverage of the food insecurity topics (i.e., Figure~\ref{fig:weekly_trends_misc_topics}(d). \textit{Food availability} dominates over \textit{food accessibility} and \textit{adequacy} in most weeks. However, \textit{food acceptability}, other than a few spikes in February and May 2020, remains less of a concern for the public, thus not discussed on Twitter. Food shortage was one of the critical issues faced by many countries around the world. This Twitter data might help detect hot-spots with severe food shortages ultimately helping authorities focus on most vulnerable areas. 

Figure~\ref{fig:weekly_trends_misc_topics}(e \& f) shows trends for mask usage and shortage as well as vaccination. The \textit{``Importance of mask''} category, which includes mask usage, importance of mask, etc., leads the discussion throughout. The \textit{mask shortage} category spikes in the early months of 2020 and then averages out. \textit{Mask violations} seem to surge in May and November 2020 and for the rest it stays steady. Mask shortage tweets worth further analysis to find out areas with severe shortages. The discussion on vaccines is comparatively lower than all other topics. However, the category on willingness to take or already taken vaccine is hopeful and spiked for the most months, in particular, late 2020 and early 2021.  

\section*{Data Records}
\label{sec:data_records}
The TBCOV dataset is shared through the CrisisNLP repository\footnote{\url{https://crisisnlp.qcri.org/tbcov}}. The dataset contains three types of releases covering different dimensions of the data. Specifically, we offer a base release including a comprehensive set of attributes such as tweet ids, user ids, sentiment labels, named-entities, geotagging results, user types, and gender labels, among others. The base release contains tab-separated values (TSV) files representing the data collection months (i.e., February \nth{1}, 2020 to March \nth{31}, 2021). 
In addition to the base data, we offer two additional releases consisting of tweet ids for the top 20 languages and top ten countries. 
The purpose of id-based releases is to maximize data accessibility for data analysts targeting one or few languages or counties for their analyses. Additional releases will be provided based on end-user demands.
We make the dataset publicly available for research and non-profit uses. Adhering to Twitter data redistribution policies, we cannot share full tweet content. 

\section*{Technical Validation}
\label{sec:validation}

\subsection*{Validation of geotagging approach}
\label{sec:geotagging_valiation}
To evaluate the proposed geotagging method we first obtain ground-truth data for different attributes. Geotagged tweets with GPS coordinates, i.e., \textit{latitude} and \textit{longitude}, were used as ground truth for the evaluation of the \textit{place} field. Specifically, tweets with \textit{(i)} \textit{geo-coordinates} and \textit{(ii)} \textit{place} fields are sampled and their location granularities such as country, state, county, and city were obtained. Finally, we compute the precision metric, i.e., the ratio of correctly predicted location granularity to the total predicted outcomes (i.e., sum of true positives and false positives). Table~\ref{tab:geo_place_evaluation} shows the evaluation results along with the number of sampled tweets (in parenthesis). All location granularity scores, except county, are promising.

The evaluation of the \textit{user location} geotagging method is performed on a manually annotated\footnote{The authors of this paper performed the manual annotation.} random sample of 500 user locations. Specifically, each user location string was examined to determine its corresponding country, state, county, and city. Google search, Wikipedia and other sources were allowed to search and disambiguate in case multiple candidates emerge. Location strings such as ``Planet earth'', were annotated as ``NA'' and used in the evaluation procedure (i.e., the system's output for an ``NA'' case is considered \textit{True Positive} if blank and \textit{False Positive} otherwise).  Table~\ref{tab:geo_userloc_evaluation} shows the evaluation results in terms of precision, recall, and F1-score. Overall, the F1-scores for all location granularities are high. However, fine-grained location resolution poses more challenges for the method (e.g., the recall at the city level is 0.656 compared to the recall of 1.0 at the country level). 

Lastly, to evaluate text-based attributes (i.e., \textit{tweet text} and \textit{user profile description}), 1,000 tweets were randomly sampled and crowdsourced on Appen\footnote{\url{https://appen.com/}}, which is a paid crowdsourcing platform. Specifically, given a tweet text, annotators were asked to \textit{(i)} tag toponyms (i.e., location names such as USA, Paris) and \textit{(ii)} specify the location type (i.e., country, state, county, and city) of the identified toponyms. Three evaluation metrics, i.e., precision, recall, and F1-score were computed using the annotated location tokens. Table~\ref{tab:geo_text_evaluation} presents geotagging evaluation results for the two text-based attributes (i.e., \textit{tweet text} and \textit{user profile description}). Geotagging at country and state levels yields promising F1-scores (i.e., 0.803 and 0.703, respectively). However, the results for county and city are weak.

\begin{table}[]
\centering
\begin{tabular}{lllll}
\toprule
 & \multicolumn{1}{c}{Country} & \multicolumn{1}{c}{State} & \multicolumn{1}{c}{County} & \multicolumn{1}{c}{City} \\
 \midrule
Place & 0.988 (7,990) & 0.967 (7,871) & 0.771 (7,394) & 0.967 (4,903)\\
\bottomrule
\end{tabular}
\caption{Geotagging method evaluation for the \textit{place} attribute (in terms of precision). Numbers in parenthesis represent the sample size.}
\label{tab:geo_place_evaluation}
\end{table}

\begin{table}[!htb]
\centering
    \begin{minipage}{.45\linewidth}
        \centering
             \begin{tabular}{lcccc}
                \toprule
                 \multicolumn{1}{l}{Metric} & \multicolumn{1}{c}{Country} & \multicolumn{1}{c}{State} & \multicolumn{1}{c}{County} & \multicolumn{1}{c}{City} \\
                 \midrule
                Precision & 0.868 & 0.839 & 0.648 & 0.802 \\
                Recall & 1.000 & 0.968 & 0.922 & 0.656 \\
                F1-score & 0.929 & 0.899 & 0.761 & 0.722\\
                \bottomrule
            \end{tabular}
            \caption{Geotagging method evaluation for the \textit{user location} attribute }
            \label{tab:geo_userloc_evaluation}
    \end{minipage}%
    \begin{minipage}{.45\linewidth}
        \centering
        \begin{tabular}{lcccc}
            \toprule
             \multicolumn{1}{l}{Metric} & \multicolumn{1}{c}{Country} & \multicolumn{1}{c}{State} & \multicolumn{1}{c}{County} & \multicolumn{1}{c}{City} \\
             \midrule
            Precision & 0.888 & 0.781 & 0.056 & 0.430 \\
            Recall & 0.732 & 0.640 & 0.462 & 0.184 \\
            F1-score & 0.803 & 0.703 & 0.100 & 0.258\\
            \bottomrule
        \end{tabular}
        \caption{Geotagging method evaluation for \textit{tweet text} \& \textit{user profile description})}
        \label{tab:geo_text_evaluation}
    \end{minipage}
\end{table}



\subsection*{Validation of person user type}
\label{sec:eval_users}

Since our main focus is on the tweets posted by the general public, here we evaluate the \textit{person} entity predictions. A random sample consisting of 200 model predictions of the \textit{person} entity is selected for the evaluation. The sampled accounts were manually checked by the authors of this paper and marked as either \textit{person} or \textit{non-person}. The manual investigation revealed 186 user accounts with correct and 14 with incorrect model predictions. This yields a precision of 0.93 for the \textit{Person} category, which is quite promising. 

\subsection*{Validation of gender classification}
\label{sec:eval_gender}
To evaluate the gender classification model, 20\% (i.e., 24,267) of the 121,335 annotated names were randomly sampled and hold out during the training phase. The unseen hold out set was used to test the model and compute several evaluation metrics. Table~\ref{tab:gender_eval} shows the evaluation results. The F1-score of the \textit{female} class is very reasonable (0.878) compared to the male class (0.807). This is probably due to the high prevalence of the female class in the training set. 

\begin{table}[!h]
\centering
  \begin{tabular}{lcccc}
    \toprule
    Metric & Female & Male & Macro avg. & Weighted avg.\\
    \midrule
    Precision & 0.872 & 0.816 &  0.844 & 0.850 \\
    Recall & 0.885 & 0.797 & 0.841 & 0.851 \\
    F1-score & 0.878 & 0.807 & 0.843 & 0.850 \\
  \bottomrule
\end{tabular}
\caption{Gender classification results (model=Random Forest)}
  \label{tab:gender_eval}
\end{table}
\section*{Usage Notes}
\label{sec:usage}


All the collected data is persisted in Elasticsearch 7.10 database. The code used for data processing is written in Python 3. The code required to hydrate tweets and to use the provided base release files is available on GitHub\footnote{\url{https://github.com/CrisisComputing/TBCOV}}. Furthermore, we postulate that this large-scale, multilingual, geotagged social media data can empower multidisciplinary research communities to perform longitudinal studies, evaluate how societies are collectively coping with this unprecedented global crisis as well as to develop computational methods to address real-world challenges, including but not limited to the following:

\begin{itemize}[noitemsep,topsep=1pt,parsep=0pt,partopsep=0pt]

\item \textbf{\textit{Disease forecasting and surveillance}} lead to the early detection and prevention of an outbreak. Moreover, early warning systems alert authorities and healthcare providers to prepare and respond to outbreaks in a timely fashion. TBCOV's broad topical coverage, particularly about self-reported symptoms and deaths, can be a strong indicator for the early warning systems. 

\item \textbf{\textit{Identification of fake information}} is essential to tackle negative influences on societies, especially during health emergencies. Tweets' temporal information, re-sharing and retweeting patterns, and the use of specific tone in the textual content can potentially lead to the identification of rumors and fake information. More than two billion tweets in the TBCOV dataset is a goldmine for detecting conspiracies, rumors, and misinformation circulated on social media (e.g., drinking bleach can cure COVID-19). More importantly, the data can be used to develop robust models for fake news and rumor detection.

\item \textbf{\textit{Understanding communities' knowledge gaps}} during emergency situations such as the COVID-19 pandemic is crucial for authorities to deal with the surge of uncertainties. TBCOV's comprehensive geographic as well as temporal coverage can be analyzed to understand public questions and queries. 

\item \textbf{\textit{Identification of shortages}} of important items such as Personal Protective Equipment (PPE), oxygen, and face mask becomes the top priority for governments during health emergencies. Building models to identify pertinent social media reports could help authorities plan and prevent devastating consequences of shortages. 

\item \textbf{\textit{Understanding public sentiment and reactions}} against governments policies such as lock downs, closure of businesses, as well as slow response or vaccination rate can be performed using social media data such as TBCOV. 

\item \textbf{\textit{Rapid needs assessment}} informs humanitarian organizations' and governments' response operations and determines relief priorities for an affected population during emergencies such as the COVID-19 pandemic. Our trends analysis results highlighted the effectiveness of TBCOV for mining priority needs of population in terms of food, cash, medicines, and more.

\item \textbf{\textit{Identification of self-reported symptoms}} such as fever, cough, loss of taste, etc. through social media data could indicate a likely future hot-spot when reports spike in a geographical area. TBCOV tweets geotagged with fine-grained locations, such as counties and cities, can be useful to build models for symptom detection and hot-spot prediction.

\item \textbf{\textit{Finding correlations}} is an important measure of relationship between two variables. We remark that the TBCOV dataset can be used to perform various types of correlation analysis to detect patterns and generate hypotheses. These analyses include, but are not limited to, finding correlations between COVID-19 cases and self-reported symptoms on Twitter; or between COVID-19 cases and death reports. Correlations between COVID-19 cases and negative sentiment in a geographical location or the surge of messages showing anxiety and unemployment rate; or correlation between daily negative tweets and the rate of food insufficiency in an area can open new avenues for interesting analyses.
\end{itemize}

The aforementioned topics mainly cover real-world applications of the TBCOV dataset. However, we believe that the dataset is useful for several computing problems such as unsupervised learning to identify clusters of related messages, transfer learning between topical domains and language domains, geographic information systems, automatic recognition and disambiguation of location mentions, named-entity extraction, topic evolution and concept-drift detection, among others.

\section*{Code availability}
\label{sec:code}
The code to use this dataset is available through \url{https://github.com/CrisisComputing/TBCOV}. The code repository contains scripts to perform hydration of tweets using the released tweet ids. The hydration process fetches full tweet content from Twitter APIs. Moreover, we provide code to use the base release data files in a more efficient way, particularly for analyses focusing on specific languages or countries.

\bibliography{bibliography}

\begin{thebibliography}{10}
\urlstyle{rm}
\expandafter\ifx\csname url\endcsname\relax
  \def\url#1{\texttt{#1}}\fi
\expandafter\ifx\csname urlprefix\endcsname\relax\def\urlprefix{URL }\fi
\expandafter\ifx\csname doiprefix\endcsname\relax\def\doiprefix{DOI: }\fi
\providecommand{\bibinfo}[2]{#2}
\providecommand{\eprint}[2][]{\url{#2}}

\bibitem{castillo2016big}
\bibinfo{author}{Castillo, C.}
\newblock \emph{\bibinfo{title}{Big Crisis Data}}
  (\bibinfo{publisher}{Cambridge University Press}, \bibinfo{year}{2016}).

\bibitem{fraustino2017social}
\bibinfo{author}{Fraustino, J.~D.}, \bibinfo{author}{Liu, B.~F.} \&
  \bibinfo{author}{Jin, Y.}
\newblock \bibinfo{journal}{\bibinfo{title}{Social media use during
  disasters}}.
\newblock {\emph{\JournalTitle{Social media and crisis communication}}}
  \textbf{\bibinfo{volume}{283}}, \bibinfo{pages}{32--47}
  (\bibinfo{year}{2017}).

\bibitem{starbird2010chatter}
\bibinfo{author}{Starbird, K.}, \bibinfo{author}{Palen, L.},
  \bibinfo{author}{Hughes, A.~L.} \& \bibinfo{author}{Vieweg, S.}
\newblock \bibinfo{title}{Chatter on the red: what hazards threat reveals about
  the social life of microblogged information}.
\newblock In \emph{\bibinfo{booktitle}{ACM Conference on Computer Supported
  Cooperative Work}}, \bibinfo{pages}{241--250} (\bibinfo{year}{2010}).

\bibitem{sinnenberg2017twitter}
\bibinfo{author}{Sinnenberg, L.} \emph{et~al.}
\newblock \bibinfo{journal}{\bibinfo{title}{Twitter as a tool for health
  research: a systematic review}}.
\newblock {\emph{\JournalTitle{American journal of public health}}}
  \textbf{\bibinfo{volume}{107}}, \bibinfo{pages}{e1--e8}
  (\bibinfo{year}{2017}).

\bibitem{zadeh2019social}
\bibinfo{author}{Zadeh, A.~H.}, \bibinfo{author}{Zolbanin, H.~M.},
  \bibinfo{author}{Sharda, R.} \& \bibinfo{author}{Delen, D.}
\newblock \bibinfo{journal}{\bibinfo{title}{Social media for nowcasting flu
  activity: Spatio-temporal big data analysis}}.
\newblock {\emph{\JournalTitle{Information Systems Frontiers}}}
  \textbf{\bibinfo{volume}{21}}, \bibinfo{pages}{743--760}
  (\bibinfo{year}{2019}).

\bibitem{broniatowski2013national}
\bibinfo{author}{Broniatowski, D.~A.}, \bibinfo{author}{Paul, M.~J.} \&
  \bibinfo{author}{Dredze, M.}
\newblock \bibinfo{journal}{\bibinfo{title}{National and local influenza
  surveillance through twitter: an analysis of the 2012-2013 influenza
  epidemic}}.
\newblock {\emph{\JournalTitle{PloS one}}} \textbf{\bibinfo{volume}{8}}
  (\bibinfo{year}{2013}).

\bibitem{lamsal2020coronasentiment}
\bibinfo{author}{Lamsal, R.}
\newblock \bibinfo{title}{{Corona Virus (COVID-19) Geolocation-based Sentiment
  Data. IEEE Dataport.}}
\newblock \bibinfo{howpublished}{\url{http://dx.doi.org/10.21227/fpsb-jz61}},
  \url{10.21227/fpsb-jz61} (\bibinfo{year}{2020}).
\newblock \bibinfo{note}{(Accessed: 2020-05-06)}.

\bibitem{lamsal2020coronatweets}
\bibinfo{author}{Lamsal, R.}
\newblock \bibinfo{title}{{Corona Virus (COVID-19) Tweets Dataset. IEEE
  Dataport.}}
\newblock \bibinfo{howpublished}{\url{http://dx.doi.org/10.21227/781w-ef42}},
  \url{10.21227/781w-ef42} (\bibinfo{year}{2020}).
\newblock \bibinfo{note}{(Accessed: 2020-05-06)}.

\bibitem{alqurashi2020large}
\bibinfo{author}{Alqurashi, S.}, \bibinfo{author}{Alhindi, A.} \&
  \bibinfo{author}{Alanazi, E.}
\newblock \bibinfo{journal}{\bibinfo{title}{{Large Arabic Twitter Dataset on
  COVID-19}}}.
\newblock {\emph{\JournalTitle{arXiv preprint arXiv:2004.04315}}}
  (\bibinfo{year}{2020}).

\bibitem{haouari2020arcov}
\bibinfo{author}{Haouari, F.}, \bibinfo{author}{Hasanain, M.},
  \bibinfo{author}{Suwaileh, R.} \& \bibinfo{author}{Elsayed, T.}
\newblock \bibinfo{journal}{\bibinfo{title}{{ArCOV-19: The First Arabic
  COVID-19 Twitter Dataset with Propagation Networks}}}.
\newblock {\emph{\JournalTitle{arXiv preprint arXiv:2004.05861}}}
  (\bibinfo{year}{2020}).

\bibitem{kang2020multiscale}
\bibinfo{author}{Kang, Y.} \emph{et~al.}
\newblock \bibinfo{journal}{\bibinfo{title}{Multiscale dynamic human mobility
  flow dataset in the us during the covid-19 epidemic}}.
\newblock {\emph{\JournalTitle{Scientific data}}} \textbf{\bibinfo{volume}{7}},
  \bibinfo{pages}{1--13} (\bibinfo{year}{2020}).

\bibitem{park2021covid}
\bibinfo{author}{Park, S.} \emph{et~al.}
\newblock \bibinfo{journal}{\bibinfo{title}{Covid-19 discourse on twitter in
  four asian countries: Case study of risk communication}}.
\newblock {\emph{\JournalTitle{Journal of medical Internet research}}}
  \textbf{\bibinfo{volume}{23}}, \bibinfo{pages}{e23272}
  (\bibinfo{year}{2021}).

\bibitem{banda2020large}
\bibinfo{author}{Banda, J.~M.} \emph{et~al.}
\newblock \bibinfo{journal}{\bibinfo{title}{{A large-scale COVID-19 Twitter
  chatter dataset for open scientific research--an international
  collaboration}}}.
\newblock {\emph{\JournalTitle{arXiv preprint arXiv:2004.03688}}}
  (\bibinfo{year}{2020}).

\bibitem{gohil2018sentiment}
\bibinfo{author}{Gohil, S.}, \bibinfo{author}{Vuik, S.} \&
  \bibinfo{author}{Darzi, A.}
\newblock \bibinfo{journal}{\bibinfo{title}{Sentiment analysis of health care
  tweets: review of the methods used}}.
\newblock {\emph{\JournalTitle{JMIR public health and surveillance}}}
  \textbf{\bibinfo{volume}{4}}, \bibinfo{pages}{e43} (\bibinfo{year}{2018}).

\bibitem{gui2017managing}
\bibinfo{author}{Gui, X.}, \bibinfo{author}{Kou, Y.}, \bibinfo{author}{Pine,
  K.~H.} \& \bibinfo{author}{Chen, Y.}
\newblock \bibinfo{title}{Managing uncertainty: using social media for risk
  assessment during a public health crisis}.
\newblock In \emph{\bibinfo{booktitle}{Proceedings of the 2017 CHI conference
  on human factors in computing systems}}, \bibinfo{pages}{4520--4533}
  (\bibinfo{year}{2017}).

\bibitem{alamoodi2020sentiment}
\bibinfo{author}{Alamoodi, A.} \emph{et~al.}
\newblock \bibinfo{journal}{\bibinfo{title}{Sentiment analysis and its
  applications in fighting covid-19 and infectious diseases: A systematic
  review}}.
\newblock {\emph{\JournalTitle{Expert systems with applications}}}
  \bibinfo{pages}{114155} (\bibinfo{year}{2020}).

\bibitem{barbieri2021xlmtwitter}
\bibinfo{author}{Barbieri, F.}, \bibinfo{author}{Espinosa-Anke, L.} \&
  \bibinfo{author}{Camacho-Collados, J.}
\newblock \bibinfo{title}{{A Multilingual Language Model Toolkit for Twitter}}.
\newblock In \emph{\bibinfo{booktitle}{arXiv preprint arXiv:2104.12250}}
  (\bibinfo{year}{2021}).

\bibitem{geotagging}
\bibinfo{title}{Geotagging}.
\newblock
  \bibinfo{howpublished}{\url{https://en.wikipedia.org/wiki/Geotagging}}.
\newblock \bibinfo{note}{Accessed: 2021-06-20}.

\bibitem{boulos2020geographical}
\bibinfo{author}{Boulos, M. N.~K.} \& \bibinfo{author}{Geraghty, E.~M.}
\newblock \bibinfo{title}{Geographical tracking and mapping of coronavirus
  disease covid-19/severe acute respiratory syndrome coronavirus 2 (sars-cov-2)
  epidemic and associated events around the world: how 21st century gis
  technologies are supporting the global fight against outbreaks and epidemics}
  (\bibinfo{year}{2020}).

\bibitem{haworth2016emergency}
\bibinfo{author}{Haworth, B.}
\newblock \bibinfo{journal}{\bibinfo{title}{Emergency management perspectives
  on volunteered geographic information: Opportunities, challenges and
  change}}.
\newblock {\emph{\JournalTitle{Computers, Environment and Urban Systems}}}
  \textbf{\bibinfo{volume}{57}}, \bibinfo{pages}{189--198}
  (\bibinfo{year}{2016}).

\bibitem{tzavella2018opportunities}
\bibinfo{author}{Tzavella, K.}, \bibinfo{author}{Fekete, A.} \&
  \bibinfo{author}{Fiedrich, F.}
\newblock \bibinfo{journal}{\bibinfo{title}{Opportunities provided by
  geographic information systems and volunteered geographic information for a
  timely emergency response during flood events in cologne, germany}}.
\newblock {\emph{\JournalTitle{Natural Hazards}}}
  \textbf{\bibinfo{volume}{91}}, \bibinfo{pages}{29--57}
  (\bibinfo{year}{2018}).

\bibitem{marrero2013named}
\bibinfo{author}{Marrero, M.}, \bibinfo{author}{Urbano, J.},
  \bibinfo{author}{S{\'a}nchez-Cuadrado, S.}, \bibinfo{author}{Morato, J.} \&
  \bibinfo{author}{G{\'o}mez-Berb{\'\i}s, J.~M.}
\newblock \bibinfo{journal}{\bibinfo{title}{Named entity recognition:
  fallacies, challenges and opportunities}}.
\newblock {\emph{\JournalTitle{Computer Standards \& Interfaces}}}
  \textbf{\bibinfo{volume}{35}}, \bibinfo{pages}{482--489}
  (\bibinfo{year}{2013}).

\bibitem{sekine2009named}
\bibinfo{author}{Sekine, S.} \& \bibinfo{author}{Ranchhod, E.}
\newblock \emph{\bibinfo{title}{Named entities: recognition, classification and
  use}}, vol.~\bibinfo{volume}{19} (\bibinfo{publisher}{John Benjamins
  Publishing}, \bibinfo{year}{2009}).

\bibitem{farmakiotou2000rule}
\bibinfo{author}{Farmakiotou, D.} \emph{et~al.}
\newblock \bibinfo{title}{Rule-based named entity recognition for greek
  financial texts}.
\newblock In \emph{\bibinfo{booktitle}{Proceedings of the Workshop on
  Computational lexicography and Multimedia Dictionaries (COMLEX 2000)}},
  \bibinfo{pages}{75--78} (\bibinfo{organization}{Citeseer},
  \bibinfo{year}{2000}).

\bibitem{finkel2009nested}
\bibinfo{author}{Finkel, J.~R.} \& \bibinfo{author}{Manning, C.~D.}
\newblock \bibinfo{title}{Nested named entity recognition}.
\newblock In \emph{\bibinfo{booktitle}{Proceedings of the 2009 conference on
  empirical methods in natural language processing}}, \bibinfo{pages}{141--150}
  (\bibinfo{year}{2009}).

\bibitem{manierre2015gaps}
\bibinfo{author}{Manierre, M.~J.}
\newblock \bibinfo{journal}{\bibinfo{title}{Gaps in knowledge: tracking and
  explaining gender differences in health information seeking}}.
\newblock {\emph{\JournalTitle{Social Science \& Medicine}}}
  \textbf{\bibinfo{volume}{128}}, \bibinfo{pages}{151--158}
  (\bibinfo{year}{2015}).

\bibitem{antonio2014gender}
\bibinfo{author}{Antonio, A.} \& \bibinfo{author}{Tuffley, D.}
\newblock \bibinfo{journal}{\bibinfo{title}{The gender digital divide in
  developing countries}}.
\newblock {\emph{\JournalTitle{Future Internet}}} \textbf{\bibinfo{volume}{6}},
  \bibinfo{pages}{673--687} (\bibinfo{year}{2014}).

\bibitem{johnson2009better}
\bibinfo{author}{Johnson, J.~L.}, \bibinfo{author}{Greaves, L.} \&
  \bibinfo{author}{Repta, R.}
\newblock \bibinfo{journal}{\bibinfo{title}{Better science with sex and gender:
  facilitating the use of a sex and gender-based analysis in health research}}.
\newblock {\emph{\JournalTitle{International journal for equity in health}}}
  \textbf{\bibinfo{volume}{8}}, \bibinfo{pages}{1--11} (\bibinfo{year}{2009}).

\bibitem{lawrence2007methodologic}
\bibinfo{author}{Lawrence, K.} \& \bibinfo{author}{Rieder, A.}
\newblock \bibinfo{journal}{\bibinfo{title}{Methodologic and ethical
  ramifications of sex and gender differences in public health research}}.
\newblock {\emph{\JournalTitle{Gender medicine}}} \textbf{\bibinfo{volume}{4}},
  \bibinfo{pages}{S96--S105} (\bibinfo{year}{2007}).

\bibitem{thara2018code}
\bibinfo{author}{Thara, S.} \& \bibinfo{author}{Poornachandran, P.}
\newblock \bibinfo{title}{Code-mixing: A brief survey}.
\newblock In \emph{\bibinfo{booktitle}{2018 International Conference on
  Advances in Computing, Communications and Informatics (ICACCI)}},
  \bibinfo{pages}{2382--2388} (\bibinfo{organization}{IEEE},
  \bibinfo{year}{2018}).

\bibitem{qazi2020geocov19}
\bibinfo{author}{Qazi, U.}, \bibinfo{author}{Imran, M.} \&
  \bibinfo{author}{Ofli, F.}
\newblock \bibinfo{journal}{\bibinfo{title}{Geocov19: a dataset of hundreds of
  millions of multilingual covid-19 tweets with location information}}.
\newblock {\emph{\JournalTitle{SIGSPATIAL Special}}}
  \textbf{\bibinfo{volume}{12}}, \bibinfo{pages}{6--15} (\bibinfo{year}{2020}).

\bibitem{mackinlay2017detection}
\bibinfo{author}{MacKinlay, A.}, \bibinfo{author}{Aamer, H.} \&
  \bibinfo{author}{Yepes, A.~J.}
\newblock \bibinfo{title}{Detection of adverse drug reactions using medical
  named entities on twitter}.
\newblock In \emph{\bibinfo{booktitle}{AMIA Annual Symposium Proceedings}},
  vol. \bibinfo{volume}{2017}, \bibinfo{pages}{1215}
  (\bibinfo{organization}{American Medical Informatics Association},
  \bibinfo{year}{2017}).

\bibitem{stefanidis2017zika}
\bibinfo{author}{Stefanidis, A.} \emph{et~al.}
\newblock \bibinfo{journal}{\bibinfo{title}{Zika in twitter: temporal
  variations of locations, actors, and concepts}}.
\newblock {\emph{\JournalTitle{JMIR public health and surveillance}}}
  \textbf{\bibinfo{volume}{3}}, \bibinfo{pages}{e22} (\bibinfo{year}{2017}).

\bibitem{li2020survey}
\bibinfo{author}{Li, J.}, \bibinfo{author}{Sun, A.}, \bibinfo{author}{Han, J.}
  \& \bibinfo{author}{Li, C.}
\newblock \bibinfo{journal}{\bibinfo{title}{A survey on deep learning for named
  entity recognition}}.
\newblock {\emph{\JournalTitle{IEEE Transactions on Knowledge and Data
  Engineering}}}  (\bibinfo{year}{2020}).

\bibitem{GRACE2021101923}
\bibinfo{author}{Grace, R.}
\newblock \bibinfo{journal}{\bibinfo{title}{Toponym usage in social media in
  emergencies}}.
\newblock {\emph{\JournalTitle{International Journal of Disaster Risk
  Reduction}}} \textbf{\bibinfo{volume}{52}}, \bibinfo{pages}{101923},
  \url{https://doi.org/10.1016/j.ijdrr.2020.101923} (\bibinfo{year}{2021}).

\bibitem{zade2018situational}
\bibinfo{author}{Zade, H.} \emph{et~al.}
\newblock \bibinfo{journal}{\bibinfo{title}{From situational awareness to
  actionability: Towards improving the utility of social media data for crisis
  response}}.
\newblock {\emph{\JournalTitle{Proceedings of the ACM on Human-Computer
  Interaction}}} \textbf{\bibinfo{volume}{2}}, \bibinfo{pages}{195}
  (\bibinfo{year}{2018}).

\bibitem{hindustan_times_helpline_21}
\bibinfo{author}{{Hindustan Times}}.
\newblock \bibinfo{title}{Inundated, covid-19 helplines crumble}.
\newblock
  \bibinfo{howpublished}{\url{https://www.hindustantimes.com/india-news/inundated-covid-helplines-crumble-101618684641863.html}}
  (\bibinfo{year}{2021}).

\bibitem{timesofindia_sm_use21}
\bibinfo{author}{{Times of India}}.
\newblock \bibinfo{title}{Social media is the new helpline}.
\newblock
  \bibinfo{howpublished}{\url{https://timesofindia.indiatimes.com/viral-news/covid-19-india-social-media-is-the-new-helpline-for-a-crisis-hit-country/articleshow/82345645.cms}}
  (\bibinfo{year}{2021}).

\bibitem{sloan2015tweets}
\bibinfo{author}{Sloan, L.}, \bibinfo{author}{Morgan, J.},
  \bibinfo{author}{Burnap, P.} \& \bibinfo{author}{Williams, M.}
\newblock \bibinfo{journal}{\bibinfo{title}{Who tweets? deriving the
  demographic characteristics of age, occupation and social class from twitter
  user meta-data}}.
\newblock {\emph{\JournalTitle{PloS one}}} \textbf{\bibinfo{volume}{10}},
  \bibinfo{pages}{e0115545} (\bibinfo{year}{2015}).

\bibitem{ajao2015survey}
\bibinfo{author}{Ajao, O.}, \bibinfo{author}{Hong, J.} \& \bibinfo{author}{Liu,
  W.}
\newblock \bibinfo{journal}{\bibinfo{title}{A survey of location inference
  techniques on twitter}}.
\newblock {\emph{\JournalTitle{Journal of Information Science}}}
  \textbf{\bibinfo{volume}{41}}, \bibinfo{pages}{855--864}
  (\bibinfo{year}{2015}).

\bibitem{carley2016crowd}
\bibinfo{author}{Carley, K.~M.}, \bibinfo{author}{Malik, M.},
  \bibinfo{author}{Landwehr, P.~M.}, \bibinfo{author}{Pfeffer, J.} \&
  \bibinfo{author}{Kowalchuck, M.}
\newblock \bibinfo{journal}{\bibinfo{title}{Crowd sourcing disaster management:
  The complex nature of twitter usage in padang indonesia}}.
\newblock {\emph{\JournalTitle{Safety science}}} \textbf{\bibinfo{volume}{90}},
  \bibinfo{pages}{48--61} (\bibinfo{year}{2016}).

\bibitem{huang2021impact}
\bibinfo{author}{Huang, H.} \emph{et~al.}
\newblock \bibinfo{journal}{\bibinfo{title}{The impact of individual behaviors
  and governmental guidance measures on pandemic-triggered public sentiment:
  Based on system dynamics and cross-validation}}.
\newblock {\emph{\JournalTitle{International journal of environmental research
  and public health}}} \textbf{\bibinfo{volume}{18}}, \bibinfo{pages}{4245}
  (\bibinfo{year}{2021}).

\bibitem{zhang2021temporal}
\bibinfo{author}{Zhang, T.} \& \bibinfo{author}{Cheng, C.}
\newblock \bibinfo{journal}{\bibinfo{title}{Temporal and spatial evolution and
  influencing factors of public sentiment in natural disasters—a case study
  of typhoon haiyan}}.
\newblock {\emph{\JournalTitle{ISPRS International Journal of
  Geo-Information}}} \textbf{\bibinfo{volume}{10}}, \bibinfo{pages}{299}
  (\bibinfo{year}{2021}).

\bibitem{o2010tweets}
\bibinfo{author}{O'Connor, B.}, \bibinfo{author}{Balasubramanyan, R.},
  \bibinfo{author}{Routledge, B.} \& \bibinfo{author}{Smith, N.}
\newblock \bibinfo{title}{From tweets to polls: Linking text sentiment to
  public opinion time series}.
\newblock In \emph{\bibinfo{booktitle}{Proceedings of the International AAAI
  Conference on Web and Social Media}}, vol.~\bibinfo{volume}{4}
  (\bibinfo{year}{2010}).

\bibitem{burnap2015cyber}
\bibinfo{author}{Burnap, P.} \& \bibinfo{author}{Williams, M.~L.}
\newblock \bibinfo{journal}{\bibinfo{title}{Cyber hate speech on twitter: An
  application of machine classification and statistical modeling for policy and
  decision making}}.
\newblock {\emph{\JournalTitle{Policy \& internet}}}
  \textbf{\bibinfo{volume}{7}}, \bibinfo{pages}{223--242}
  (\bibinfo{year}{2015}).

\bibitem{beigi2016overview}
\bibinfo{author}{Beigi, G.}, \bibinfo{author}{Hu, X.},
  \bibinfo{author}{Maciejewski, R.} \& \bibinfo{author}{Liu, H.}
\newblock \bibinfo{journal}{\bibinfo{title}{An overview of sentiment analysis
  in social media and its applications in disaster relief}}.
\newblock {\emph{\JournalTitle{Sentiment analysis and ontology engineering}}}
  \bibinfo{pages}{313--340} (\bibinfo{year}{2016}).

\bibitem{aday2012new}
\bibinfo{author}{Aday, S.}, \bibinfo{author}{Farrell, H.},
  \bibinfo{author}{Lynch, M.}, \bibinfo{author}{Sides, J.} \&
  \bibinfo{author}{Freelon, D.}
\newblock \bibinfo{journal}{\bibinfo{title}{New media and conflict after the
  arab spring}}.
\newblock {\emph{\JournalTitle{United States Institute of Peace}}}
  \textbf{\bibinfo{volume}{80}}, \bibinfo{pages}{1--24} (\bibinfo{year}{2012}).

\bibitem{liu2012sentiment}
\bibinfo{author}{Liu, B.}
\newblock \bibinfo{journal}{\bibinfo{title}{Sentiment analysis and opinion
  mining}}.
\newblock {\emph{\JournalTitle{Synthesis lectures on human language
  technologies}}} \textbf{\bibinfo{volume}{5}}, \bibinfo{pages}{1--167}
  (\bibinfo{year}{2012}).

\bibitem{medhat2014sentiment}
\bibinfo{author}{Medhat, W.}, \bibinfo{author}{Hassan, A.} \&
  \bibinfo{author}{Korashy, H.}
\newblock \bibinfo{journal}{\bibinfo{title}{Sentiment analysis algorithms and
  applications: A survey}}.
\newblock {\emph{\JournalTitle{Ain Shams engineering journal}}}
  \textbf{\bibinfo{volume}{5}}, \bibinfo{pages}{1093--1113}
  (\bibinfo{year}{2014}).

\bibitem{zhang2018deep}
\bibinfo{author}{Zhang, L.}, \bibinfo{author}{Wang, S.} \&
  \bibinfo{author}{Liu, B.}
\newblock \bibinfo{journal}{\bibinfo{title}{Deep learning for sentiment
  analysis: A survey}}.
\newblock {\emph{\JournalTitle{Wiley Interdisciplinary Reviews: Data Mining and
  Knowledge Discovery}}} \textbf{\bibinfo{volume}{8}}, \bibinfo{pages}{e1253}
  (\bibinfo{year}{2018}).

\bibitem{yue2019survey}
\bibinfo{author}{Yue, L.}, \bibinfo{author}{Chen, W.}, \bibinfo{author}{Li,
  X.}, \bibinfo{author}{Zuo, W.} \& \bibinfo{author}{Yin, M.}
\newblock \bibinfo{journal}{\bibinfo{title}{A survey of sentiment analysis in
  social media}}.
\newblock {\emph{\JournalTitle{Knowledge and Information Systems}}}
  \textbf{\bibinfo{volume}{60}}, \bibinfo{pages}{617--663}
  (\bibinfo{year}{2019}).

\bibitem{ceron2014every}
\bibinfo{author}{Ceron, A.}, \bibinfo{author}{Curini, L.},
  \bibinfo{author}{Iacus, S.~M.} \& \bibinfo{author}{Porro, G.}
\newblock \bibinfo{journal}{\bibinfo{title}{Every tweet counts? how sentiment
  analysis of social media can improve our knowledge of citizens’ political
  preferences with an application to italy and france}}.
\newblock {\emph{\JournalTitle{New media \& society}}}
  \textbf{\bibinfo{volume}{16}}, \bibinfo{pages}{340--358}
  (\bibinfo{year}{2014}).

\bibitem{conneau2020unsupervised}
\bibinfo{author}{Conneau, A.} \emph{et~al.}
\newblock \bibinfo{title}{Unsupervised cross-lingual representation learning at
  scale}.
\newblock In \emph{\bibinfo{booktitle}{Proceedings of the 58th Annual Meeting
  of the Association for Computational Linguistics}},
  \bibinfo{pages}{8440--8451} (\bibinfo{year}{2020}).

\bibitem{twitter_stats}
\bibinfo{title}{Twitter statistics}.
\newblock
  \bibinfo{howpublished}{\url{https://www.businessofapps.com/data/twitter-statistics/}}.
\newblock \bibinfo{note}{Accessed: 2021-06-22}.

\bibitem{zhang2019less}
\bibinfo{author}{Zhang, Z.} \& \bibinfo{author}{Bors, G.}
\newblock \bibinfo{journal}{\bibinfo{title}{“less is more”: Mining useful
  features from twitter user profiles for twitter user classification in the
  public health domain}}.
\newblock {\emph{\JournalTitle{Online Information Review}}}
  (\bibinfo{year}{2019}).

\bibitem{uddin2014understanding}
\bibinfo{author}{Uddin, M.~M.}, \bibinfo{author}{Imran, M.} \&
  \bibinfo{author}{Sajjad, H.}
\newblock \bibinfo{journal}{\bibinfo{title}{Understanding types of users on
  twitter}}.
\newblock {\emph{\JournalTitle{arXiv preprint arXiv:1406.1335}}}
  (\bibinfo{year}{2014}).

\bibitem{okazaki2015using}
\bibinfo{author}{Okazaki, S.}, \bibinfo{author}{D{\'\i}az-Mart{\'\i}n, A.~M.},
  \bibinfo{author}{Rozano, M.} \& \bibinfo{author}{Men{\'e}ndez-Benito, H.~D.}
\newblock \bibinfo{journal}{\bibinfo{title}{Using twitter to engage with
  customers: a data mining approach}}.
\newblock {\emph{\JournalTitle{Internet Research}}}  (\bibinfo{year}{2015}).

\bibitem{hannon2010recommending}
\bibinfo{author}{Hannon, J.}, \bibinfo{author}{Bennett, M.} \&
  \bibinfo{author}{Smyth, B.}
\newblock \bibinfo{title}{Recommending twitter users to follow using content
  and collaborative filtering approaches}.
\newblock In \emph{\bibinfo{booktitle}{Proceedings of the fourth ACM conference
  on Recommender systems}}, \bibinfo{pages}{199--206} (\bibinfo{year}{2010}).

\bibitem{garcia2013catstream}
\bibinfo{author}{Garcia~Esparza, S.}, \bibinfo{author}{O'Mahony, M.~P.} \&
  \bibinfo{author}{Smyth, B.}
\newblock \bibinfo{title}{Catstream: categorising tweets for user profiling and
  stream filtering}.
\newblock In \emph{\bibinfo{booktitle}{Proceedings of the 2013 international
  conference on Intelligent user interfaces}}, \bibinfo{pages}{25--36}
  (\bibinfo{year}{2013}).

\bibitem{ali2019morphological}
\bibinfo{author}{Ali, M.}
\newblock \emph{\bibinfo{title}{The Morphological Gender Assignment for English
  Personal Names}}.
\newblock Ph.D. thesis, \bibinfo{school}{CALIFORNIA STATE UNIVERSITY,
  NORTHRIDGE} (\bibinfo{year}{2019}).

\bibitem{slepian2016voiced}
\bibinfo{author}{Slepian, M.~L.} \& \bibinfo{author}{Galinsky, A.~D.}
\newblock \bibinfo{journal}{\bibinfo{title}{The voiced pronunciation of initial
  phonemes predicts the gender of names.}}
\newblock {\emph{\JournalTitle{Journal of Personality and Social Psychology}}}
  \textbf{\bibinfo{volume}{110}}, \bibinfo{pages}{509} (\bibinfo{year}{2016}).

\bibitem{dataw_gender_data}
\bibinfo{author}{Babu, A.}
\newblock \bibinfo{title}{Data world: Gender by names dataset}.
\newblock
  \bibinfo{howpublished}{\url{https://data.world/arunbabu/gender-by-names}}.
\newblock \bibinfo{note}{Accessed: 2021-01-21}.

\bibitem{cmu_gender_data}
\bibinfo{author}{Kantrowitz, M.}
\newblock \bibinfo{title}{Cmu: Names gender dataset}.
\newblock
  \bibinfo{howpublished}{\url{http://www.cs.cmu.edu/afs/cs/project/ai-repository/ai/areas/nlp/corpora/names/}}.
\newblock \bibinfo{note}{Accessed: 2021-01-21}.

\bibitem{dataw_gender_data2}
\bibinfo{author}{Howard, D.}
\newblock \bibinfo{title}{Data world: Names gender dataset 2}.
\newblock
  \bibinfo{howpublished}{\url{https://data.world/howarder/gender-by-name}}.
\newblock \bibinfo{note}{Accessed: 2021-01-21}.

\bibitem{rish2001empirical}
\bibinfo{author}{Rish, I.} \emph{et~al.}
\newblock \bibinfo{title}{An empirical study of the naive bayes classifier}.
\newblock In \emph{\bibinfo{booktitle}{IJCAI 2001 workshop on empirical methods
  in artificial intelligence}}, vol.~\bibinfo{volume}{3},
  \bibinfo{pages}{41--46} (\bibinfo{year}{2001}).

\bibitem{quinlan1986induction}
\bibinfo{author}{Quinlan, J.~R.}
\newblock \bibinfo{journal}{\bibinfo{title}{Induction of decision trees}}.
\newblock {\emph{\JournalTitle{Machine learning}}}
  \textbf{\bibinfo{volume}{1}}, \bibinfo{pages}{81--106}
  (\bibinfo{year}{1986}).

\bibitem{breiman2001random}
\bibinfo{author}{Breiman, L.}
\newblock \bibinfo{journal}{\bibinfo{title}{Random forests}}.
\newblock {\emph{\JournalTitle{Machine learning}}}
  \textbf{\bibinfo{volume}{45}}, \bibinfo{pages}{5--32} (\bibinfo{year}{2001}).

\end{thebibliography}

\end{document}